\documentclass[prb,twocolumn,showpacs,preprintnumbers,float]{revtex4-2}
\usepackage[colorlinks=true,urlcolor=blue,citecolor=blue,linkcolor=blue,breaklinks=true]{hyperref}
\usepackage{graphicx}
\usepackage{amsmath}
\usepackage{amssymb}
\usepackage{times}
\usepackage{color}
\usepackage{ulem}
\usepackage{physics}
\usepackage{cleveref}
\usepackage{xr}
\usepackage{multirow}
\usepackage{comment}

\def\a{\alpha}

\def\s{\sigma}
\def\sb{\bar{\sigma}}
\def\t{\tau}
\def\w{\omega}
\def\ua{\uparrow}
\def\da{\downarrow}

\def\Vec#1{\mathbf #1}

\def\cc{\hat{c}}
\def\hh{\hat{h}}
\def\nn{\hat{n}}
\def\qq{\Vec{q}}
\def\rr{\Vec{r}}
\newcommand{\nb}{\bar{n}}

\begin{document}

\title{Hyperuniform electron distributions controlled by electron interactions in quasicrystal}

\author{Shiro Sakai$^1$, Ryotaro Arita$^{1,2}$, and Tomi Ohtsuki$^3$}
\affiliation{
$^1$Center for Emergent Matter Science, RIKEN, Wako, Saitama 351-0198, Japan\\
$^2$Department of Applied Physics, University of Tokyo, Hongo, Tokyo 113-8656, Japan\\
$^3$Physics Division, Sophia University, Chiyoda-ku, Tokyo 102-8554, Japan
}
\date{\today}
\begin{abstract}
We study how the electron-electron interactions influence the charge distributions in the metallic state of quasicrystals.
As a simple theoretical model, we introduce an extended Hubbard model on the Penrose lattice, and numerically solve the model (up to $\sim1.4$ million sites) within the Hartree-Fock approximation.
Because each site on the quasiperiodic lattice has a different local geometry, the Coulomb interaction, in particular the intersite one, works in a site-dependent way, leading to a nontrivial redistribution of the charge.
The resultant charge distribution patterns are not multifractal but characterized by hyperuniformity, which offers a measure to distinguish various inhomogeneous but ordered distributions.
We clarify how the electron interactions alter the order metric of the hyperuniformity, revealing that the intersite interaction considerably affects the hyperuniformity in particular on the electron-rich side.
\end{abstract}
\maketitle

\section{introduction}\label{sec:intro}

Quasicrystal possesses an orderly but aperiodic arrangement of atoms, characterized by an unusual rotational symmetry and self-similarity \cite{mackay82,shechtman84,levine84}. 
Such a structure may originate novel electron states and properties distinct from those of periodic crystals, as well as disordered systems.
Theoretical studies of the electron states on related quasiperiodic lattices have indeed revealed interesting spatially inhomogeneous but orderly distributions of the wave function amplitude \cite{sutherland86,tokihiro88,arai88,mace17} and electron density \cite{sakai21,rai21} in the normal phase, according to the underlying lattice structure. 
In symmetry broken phases, further interesting spatial patterns have been reported for the magnetic moment of quasiperiodic antiferromagnets \cite{wessel03,vedmedenko04,jagannathan04,wessel05,jagannathan07,szallas09,jagannathan12,thiem15,koga17,koga20,hauck21,watanabe21} and the order parameters of superconductivity \cite{sakai17,araujo19,sakai19,cao20,zhang20,nagai20,takemori20,nagai21} and the excitonic insulator \cite{inayoshi20}.

For non-interacting electrons on quasiperiodic lattices, the spatial distribution of the wave function was intensively studied in the 1980s.
These studies revealed that the wave functions of the Fibonacci model and most of those on the Penrose lattice are critical \cite{kohmoto83,ostlund83,sutherland86,niu86,kohmoto87,ma89,tsunetsugu91PRB1}, namely neither extended nor localized.
The distributions of these wave-function amplitudes are characterized by the multifractality \cite{halsey86,sutherland87,tokihiro88,mace17,jagannathan20}.

On the other hand, it is not always true that other spatial patterns are characterized by multifractality.
In fact, the charge distribution does not generally show a multifractality, i.e., a nontrivial multifractal dimension deviated from the spatial dimension,  while it still shows an interesting, seemingly self-similar, pattern.
This nonuniform density distribution makes the electron-electron interactions work differently from site to site, so that their effect further alters the spatial distribution in a nontrivial way \cite{sakai21}.
Although this can result in distributions significantly different from the non-interacting one, this change cannot be characterized by the conventional charge-order parameter as the translational symmetry is broken in the first place.
A similar situation would occur for the magnetic moment in magnetic quasicrystals, and the order parameter in superconducting or excitonic-insulating quasicrystals. 

\begin{figure}[tb]
\center{
\includegraphics[width=0.48\textwidth]{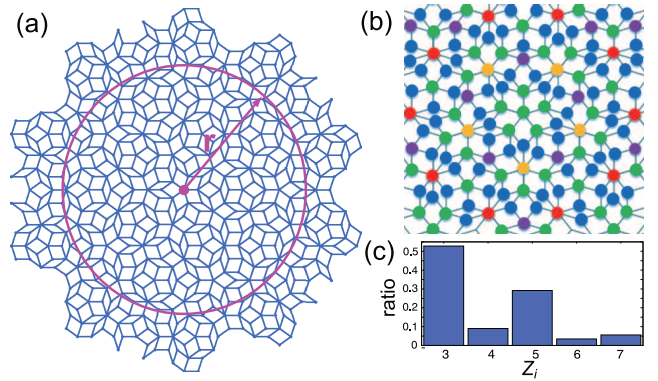}}
\caption{(a) Penrose-tiling cluster of a relatively small size ($N=601$). Magenta circle illustrates an area of radius $r$ around the center site. (b) Example of the distribution of the sites with different coordination numbers, $Z_i=3$ (blue), 4 (purple), 5 (green), 6 (yellow), and 7 (red). (c) Ratio of the sites with $Z_i$ neighbors.}
\label{fig:penrose}
\end{figure}

Thus, the variety of nonuniform spatial patterns, appearing as electron properties on quasiperiodic lattices, pose an intriguing problem, how to characterize and distinguish them.
In this paper, we apply the idea of the hyperuniformity \cite{torquato03,torquato18} to this problem.
The hyperuniformity was originally proposed by Torquato and his collaborators to quantify the distribution of a point set, and was relatively recently generalized to a random scalar field \cite{torquato16,ma17,torquato18}.
It measures a density fluctuation of a given point set (or scalar field) distributed in a $d$-dimensional space, by defining a window to count the density contained inside of the window. 
With varying the window position in the space, we obtain the variance $\sigma^2(R)$ of the density inside the window of a linear size $R$.
Then, the dependence of the variance on $R$ distinguishes various distributions, including periodic, quasiperiodic, and random ones: 
While a random distribution gives $\sigma^2(R)\propto R^d$, periodic and quasiperiodic distributions show $\sigma^2(R)\propto R^{d-1}$ for a large $R$.
The latter means that the variance is contributed only from the surface area of the window thanks to the regularity of the lattice, and such a distribution is called hyperuniform.
Moreover, within the hyperuniform distributions, the coefficient of $R^{d-1}$ measures the regularity of the distribution and is called an order metric.  

Since hyperuniformity concerns the global nature of the distribution, it can be a useful tool to characterize an electron state in quasicrystals,
particularly when it defies a conventional characterization based on symmetry, topology, or multifractality.
This is the case for the charge distributions, as we mentioned above.

In this study, we consider an extended Hubbard model on the Penrose lattice [Fig.~\ref{fig:penrose}(a)] and solve it within the Hartree-Fock approximation for a large-size cluster (up to 1~364~431 sites).
We first show that the charge distribution indeed changes significantly as the chemical potential and the strength of the electron-electron interactions are varied. In particular, the intersite interactions work electron-hole asymmetrically and have a strong influence on the distribution.
We then show that the variation of the electron density at each site can be roughly classified by the local geometry of the site while a fine structure is also discerned. 
The resultant charge distributions are analyzed in terms of the local charge variance, multifractality, and hyperuniformity.
We find that these charge distributions are not multifractal but hyperuniform, and its order metric systematically changes with the interaction strength and the average electron filling, suggesting that it offers a useful measure to quantify these inhomogeneous distributions on quasiperiodic lattices.
In addition, we also study the Aubry-Andr\'e-Harper model \cite{aubry80,harper55} as another test case of the hyperuniformity analysis, where we find that the charge distribution is always hyperuniform but its class \cite{torquato18}  changes at the self-dual point.

While the hyperuniformity of quasiperiodic point patterns has been studied in the literature \cite{zachary09,oguz17,lin17,torquato18}, here we extend its application to the electron properties on quasiperiodic lattices, to characterize their nontrivial distributions.
We expect that the hyperuniformity will be useful to characterize other electron properties, such as the magnetic moment and superconducting order parameters, in quasicrystals, too.

The remainder of the paper is organized as follows. In Sec.~\ref{sec:method}, we define the model and method, giving some intuition about the effects of electron-electron interactions. We also briefly describe the definition of the perpendicular space and multifractal dimension, and the way to calculate the order metric of the hyperuniformity. 
Section \ref{sec:results} is devoted to the calculated results.
We first clarify how the electron interactions alter the spatial distribution of the charge (Secs.~\ref{ssec:u} and \ref{ssec:v} ), finding  
various interaction-driven charge distributions (Sec.~\ref{ssec:rmap}) distinct from those induced by the electron hopping in the noninteracting system.
The perpendicular-space mapping (Sec.~\ref{ssec:perp_result}) clarifies the dependence of the charge density on the local geometry around each site.
The local density variance (Sec.~\ref{ssec:n2}) shows interesting nonmonotonic dependencies on the intersite-interaction strength and average electron filling.
In Sec.~\ref{ssec:ma}, we demonstrate that the charge distribution is {\it not} characterized by the multifractality.
In Sec.~\ref{ssec:hyper_result}, we show the calculated order metric of the hyperuniformity to characterize the various real-space patterns. 
We add several discussions in Sec.~\ref{sec:discuss}, and summarize the paper in Sec.~\ref{sec:summary}.
Appendices are devoted to the additional information on the effect of the Fock term, the momentum-space profiles, the real-space mapping of the local density of states, and a hyperuniformity analysis of the  Aubry-Andr\'e-Harper model \cite{aubry80,harper55}.

\section{Model and Method}\label{sec:method}
\subsection{Extended Hubbard model on Penrose tiling}\label{ssec:model}
The Penrose tiling is a prototypical two-dimensional structure of a quasicrystal. It covers a plane with only two types of rhombuses [Fig.~\ref{fig:penrose}(a)]. The structure is constructed deterministically by applying the inflation-deflation rule \cite{levine84} iteratively to an initially small cluster.
We regard each vertex of a rhombus as a site and consider an electron hopping through the edges of the rhombus.
We set the length of the edge as the unit of length.
As shown in Fig.~\ref{fig:penrose}(b), each site (indexed by $i$) has different coordination numbers $Z_i$, ranging from 3 to 7 \footnote{Although some sites at the boundary have $Z_i=2$, we do not incorporate these sites in any plots below.}.
Figure \ref{fig:penrose}(c) shows the ratio of the sites with $Z_i$ neighbors: We see that the $Z_i=3$ sites are dominant while the $Z_i=6$ and 7 sites are much less.
The geometry beyond the nearest neighbors also differs between sites.
This geometrical inhomogeneity leads to the inhomogeneity in the local electron properties like the electron density and the local density of states. 

We consider the extended Hubbard model on the Penrose tiling. The Hamiltonian reads 
\begin{align}
\hat{H}&=-t \sum_{\langle ij\rangle\s} (\cc_{i\s}^\dagger \cc_{j\s}^{\phantom {\dagger}}+{\it h.c.}) - \mu\sum_{i\s}\nn_{i\s}\nonumber\\
&+U\sum_{i}\nn_{i\ua} \nn_{i\da} +V\sum_{\langle ij\rangle\s\s'}\nn_{i\s} \nn_{j\s'}, 
\label{eq:hubbard}
\end{align}
where $\hat{c}_{i\s}$ ($\cc_{i\s}^\dagger$) annihilates (creates) an electron of spin $\s (=\ua, \da)$ at site $i$ and $\nn_{i\s}\equiv \cc_{i\s}^\dagger \cc_{i\s}$.
The electron hopping $t=1$ is defined between the neighboring two sites (denoted by $\langle ij\rangle$) connected by the edge of the rhombuses. We note that the bare "bandwidth" of the site-averaged density of states is about $8.5t$. 
$U$ and $V$ represent the strength of the onsite and nearest-neighbor Coulomb repulsions, respectively.
The chemical potential $\mu$ is determined self-consistently to fix the average electron density, $\nb\equiv \frac{1}{N}\sum_{i} n_{i}$ with $n_{i}\equiv\sum_{\s}\langle \nn_{i\s}\rangle$, at a given value. 
Here, $N$ is the number of sites.
Note that a peculiar electronic structure, called confined state \cite{kohmoto86PRL,arai88,day-roberts20}, is present at half filling ($\nb=1$) in the non-interacting system.
To make a general statement, we avoid these states, focusing the fillings away from the half filling.
Note also that the Penrose lattice is bipartite so that the hole doping and electron doping to $\nb=1$ are equivalent as far as $V=0$.

\subsection{Mean-field approximations}\label{ssec:mfa}
Within the Hartree-Fock approximation, the interaction part of the Hamiltonian (\ref{eq:hubbard}) is reduced to
\begin{align}
\sum_{i\s}\left[U\langle \nn_{i\sb}\rangle + V\sum_{j:{\rm n.n.\ of \ } i}n_{j} \right]  \nn_{i\s}
-V\sum_{\langle ij\rangle\s} \langle \cc_{i\s}^\dagger \cc_{j\s}^{\phantom {\dagger}}\rangle \cc_{j\s}^\dagger \cc_{i\s}^{\phantom {\dagger}}. \label{eq:mf} 
\end{align}
We see that, while the $U$ term depends on site only through the local electron density $\langle \nn_{i\sb}\rangle$, the effect of $V$ explicitly depends on the geometry around the site:
The Hartree term of $V$ gives a strongly site-dependent potential through the charge distribution at neighboring sites while the Fock term effectively alters the electron hopping in a site-dependent way. 

Moreover, we can deduce that the effect of $V$ can be different between the hole-doped ($\nb<1$) and electron-doped ($\nb>1$) sides \cite{sakai21}: 
Under the electron-hole transformation, $\cc_{i\s}^\dagger \to (-1)^i \hh_{i\s}$ and $\cc_{i\s} \to (-1)^i \hh_{i\s}^\dagger$ (where the factor $(-1)^i$ changes sign when the sublattice changes), the nearest-neighbor interaction is transformed as 
\begin{align}
V\sum_{\langle ij\rangle\s\s'} \nn_{i\s} \nn_{j\s'}\to -2V\sum_{i\s} Z_i \nn_{i\s}^h +V\sum_{\langle ij\rangle\s\s'} \nn_{i\s}^h \nn_{j\s'}^h
\end{align}
with $\nn_{i\s}^{h}\equiv \hh_{i\s}^\dagger \hh_{i\s}$.
The first term on the right-hand side gives a site-dependent potential, which cannot be absorbed into the chemical potential, unlike the case of periodic systems. This term induces an asymmetry between the hole and electron sides \cite{sakai21}. Note that, in quasiperiodic systems, a similar electron-hole asymmetry would occur for other types of intersite interactions, too.

We solve the mean-field Hamiltonian self-consistently at zero temperature through the kernel polynomial method \cite{weisse06}, where we expand the local density of states $\rho_i(\w)$ 
in terms of the Chebyshev polynomials. 
The charge density is then obtained as $n_i=\int_{-\infty}^{\mu} \rho_i(\w)d\w$.

Using the idea of the localized Krylov subspace \cite{furukawa04,nagai20}, we can calculate the coefficients of the Chebyshev-polynomial expansion efficiently.
This enables us to deal with large-size clusters (up to 1~364~431 sites in this study).  We have checked how the results depend on the order $K$ of the Chebyshev-polynomial expansion and confirmed that $K=500$ gives sufficiently accurate results for $n_i$. 

The main results presented in this paper were obtained with open-boundary clusters while we examined several different boundary conditions, open boundary, periodic boundary (with periodic approximants) \cite{tsunetsugu86,entin-wohlman88}, and the local mirror boundary \cite{kalugin14}, and confirmed that the difference is negligible in the results we present in this paper.
Then, the open-boundary cluster is advantageous to the other two because we can use the $C_{5v}$ symmetry of the cluster to reduce the computational cost of the Hartree-Fock calculation.
To plot $n_i$, we use only the inner sites away from the boundary of a sufficiently large open-boundary cluster ($N>10^4$).
Note that, unlike the wave functions, the distribution of $n_i$ is insensitive to the boundary conditions as far as we concentrate on the inner sites. 

\subsection{Perpendicular-space mapping}\label{ssec:perp}

When the Penrose tiling is constructed by projecting a five-dimensional hypercubic lattice onto a two-dimensional plane (physical space), the dimensions perpendicular to the physical space is called the perpendicular space.
The position in the perpendicular space reflects the local geometry of the site in the physical space. Namely, sites in a similar ambient geometry in the physical space are assembled into the same area in the perpendicular space. 
Thereby, the perpendicular-space map has often been used to analyze the real-space patterns on quasiperiodic lattices \cite{jagannathan07,koga17,inayoshi20,takemori20,koga20,ghadimi20,mirzhalilov20}.

Suppose $\{\Vec{d}_i\}_{i=1,\cdots,5}$ are the primitive lattice vectors of the hypercubic lattice in five dimensions so that a lattice point is expressed as $\Vec{m}^{5D}=\sum_{i=1}^5 m_i \Vec{d}_i$ with integers $\{m_i\}$.
When a lattice point on the Penrose lattice is represented as $\Vec{m}^{2D}=(x,y)=(\Vec{m}^{5D}\cdot\Vec{e}_x,\Vec{m}^{5D}\cdot\Vec{e}_y)$ by the projection, its coordinate in the perpendicular space is given by $\Vec{m}^{\perp}=(\tilde{x},\tilde{y},\tilde{z})=(\Vec{m}^{5D}\cdot\tilde{\Vec{e}}_x,\Vec{m}^{5D}\cdot\tilde{\Vec{e}}_y,\Vec{m}^{5D}\cdot\tilde{\Vec{e}}_z)$. Here, $\Vec{e}_x$, $\Vec{e}_y$, $\tilde{\Vec{e}}_x$, $\tilde{\Vec{e}}_y$, and $\tilde{\Vec{e}}_z$ are the orthonormal basis of the five-dimensinal space and the first two vectors span the physical space while the latter three span the perpendicular space.

The relation between the above vectors is given by
\begin{align}
  \begin{pmatrix}
    \Vec{d}_1\\
    \Vec{d}_2\\
    \Vec{d}_3\\
    \Vec{d}_4\\
    \Vec{d}_5
  \end{pmatrix}
  =\sqrt{\frac{2}{5}}
  \begin{pmatrix}
    c_0 & s_0 & c_0 & s_0 & \frac{1}{\sqrt{2}}\\
    c_1 & s_1 & c_2 & s_2 & \frac{1}{\sqrt{2}}\\
    c_2 & s_2 & c_4 & s_4 & \frac{1}{\sqrt{2}}\\
    c_3 & s_3 & c_6 & s_6 & \frac{1}{\sqrt{2}}\\
    c_4 & s_4 & c_8 & s_8 & \frac{1}{\sqrt{2}}
  \end{pmatrix}
  \begin{pmatrix}
    \Vec{e}_x\\
    \Vec{e}_y\\
    \tilde{\Vec{e}}_x\\
    \tilde{\Vec{e}}_y\\
    \tilde{\Vec{e}}_z
  \end{pmatrix}
\end{align}
with $c_l\equiv \cos(\frac{2\pi l}{5})$ and $s_l\equiv \sin(\frac{2\pi l}{5})$.
Since the physical space has an irrational gradient in the five-dimensional space, there is a one-to-one correspondence between the physical-space and perpendicular-space coordinates.
We can therefore calculate the perpendicular-space coordinate from a given physical-space coordinate straightforwardly.

Thanks to the translational symmetry of the hypercubic lattice, $\tilde{z}=\sum_{i=1}^5 m_i$ takes only four essentially inequivalent values $\{0,1,2,3\}$.
Because that odd and even numbers of $\tilde{z}$ correspond to different sublattices of the Penrose tiling, it is sufficient to look into only the planes of $\tilde{z}=0$ and $2$ unless the sublattice symmetry is broken.

\subsection{Multifractal dimension}\label{ssec:md}
One possible way to characterize an inhomogeneous spatial pattern is a multifractal analysis \cite{halsey86}, which has indeed been used for the wave functions on the Penrose lattice \cite{sutherland87,tokihiro88,mace17}.

For a system with a linear size $L$, we define a quantity 
\begin{align}
  ||n||_{q}^{(L)}\equiv \frac{\sum_i {n_i}^q}{(\sum_i n_i)^q}. \label{eq:norm}
\end{align}
This quantity scales as $L^{-\tau(q)}$ for a sufficiently large $L$. The exponent $\tau(q)$ is related to the multifractal dimension $D_q$ as $\tau(q)=(q-1)D_q$, i.e.,
\begin{align}
  D_q\equiv \lim_{L\to\infty} D_q^{(L)}
 \end{align}
 with
\begin{align}
  D_q^{(L)}\equiv \frac{1}{1-q}\frac{\ln ||n||_{q}^{(L)}}{\ln L}. \label{eq:dql}
 \end{align}

When $D_q$ deviates from the spatial dimension $d$ ($=2$ in the present case), the system is called multifractal. $D_q$ is generally a non-increasing function with respect to $q$ and $D_0=d$. Therefore, to judge if the system is multifractal or not, it is sufficient to examine if $D_q$ deviates from $d$ at a large $|q|$.

\subsection{Hyperuniformity}\label{ssec:hyper}
As we will see in Sec.~\ref{ssec:ma}, the calculated multifractal dimension of the charge distribution shows the trivial value (i.e., $D_q=d$). We therefore need another quantity to characterize the charge distribution. To this end, we apply the idea of the hyperuniformity introduced by Torquato and his coworkers \cite{torquato03,torquato16,torquato18}.
The hyperuniformity was originally defined for a point patterns distributed in a space, but was recently generalized to a distribution of a scalar field, too \cite{torquato16,torquato18,ma17}.

Considering a circular window of the radius $R$, we measure the variance of the electron density inside the window. 
Namely, for each center position $\Vec{r}_c$ of the window, we calculate the quantity $N(R)=\sum_{i=1}^{N} n_i \Theta(R-|\Vec{r}_i-\Vec{r}_c|)$ with the Heaviside step function $\Theta(r)$.
Then, its variance is given by
\begin{align}
  \s^2(R) = \overline{N(R)^2} - \left[ \overline{N(R)} \right]^2, \label{eq:s2}
\end{align}
where $\overline{A}\equiv \frac{1}{v}\int_v A d\Vec{r}_c$ represents the average with respect to the center position $\Vec{r}_c$ over the space of the volume $v$.
While $\s^2(R)$ is proportional to $R^d$ for a random distribution of $n_i$, the system with $\s^2(R)\propto R^{d-1}$ is called hyperuniform. 
$\s^2(R)\propto R^{d-1}$ means that the variance is contributed only from the surface area of the window, not from the volume.
Point distributions (i.e., $n_i\equiv 1$) on periodic and quasiperiodic lattices are known to be hyperuniform \cite{torquato03,torquato18}.

When we expand $\s^2(R)$ as
\begin{align}
  \s^2(R) = A R^d +B R^{d-1} +O(R^{d-2}) \label{eq:s2-2}
\end{align}
 for a large $R$,  $A$ goes to zero as $R$ goes to infinity in a hyperuniform system while $A$ remains finite in a nonhyperuniform system.
 In a hyperuniform system, $B$ is called the order metric, which represents the regularity of the hyperuniform distribution: 
 $B$ tends to be smaller for a simpler lattice \cite{torquato18}.

To evaluate $A$ and $B$ numerically, we rewrite Eq.~(\ref{eq:s2}) as
\begin{align}
  \s^2(R) &= \overline{  \left\{ \sum_{i=1}^{N} n_i \left[ \Theta(R-|\Vec{r}_i-\Vec{r}_c|)-  \frac{\pi R^2}{v}\right] \right\}^2}\nonumber\\
              &= \frac{\pi R^2}{v} \sum_{i,j=1}^{N} n_i n_j \left[ \a(|\Vec{r}_i-\Vec{r}_j|;R) - \frac{\pi R^2}{v} \right], \label{eq:s2-3}
\end{align}
where we have used $\overline{N(R)} = \frac{\pi R^2}{v}\sum_{i=1}^{N}n_i$ and $\overline{\Theta(R-|\Vec{r}_i-\Vec{r}_c|)}=\frac{\pi R^2}{v}$, and defined the scaled intersection volume, 
\begin{align}
  &\alpha (r;R) \nonumber\\
  &\equiv \frac{1}{\pi R^2}\int_v \Theta(R-|\Vec{r}_i-\Vec{r}_c|) \Theta(R-|\Vec{r}_j-\Vec{r}_c|) d\Vec{r}_c\nonumber\\
                    &=\frac{2}{\pi}\left[ \arccos(\frac{r}{2R})-\frac{r}{2R}\sqrt{1-\left(\frac{r}{2R}\right)^2} \right]\Theta(2R-r), \label{eq:alpha}
\end{align}
with $r=|\Vec{r}_i-\Vec{r}_j|$ after Refs.~\onlinecite{torquato03,torquato18}.
From Eq.~(\ref{eq:s2-3}), the coefficient $A$ in Eq.~(\ref{eq:s2-2}) is given by
\begin{align}
  A=\lim_{R\to\infty}  \frac{\pi}{v} \sum_{i,j=1}^{N} n_i n_j \left[ \a(|\Vec{r}_i-\Vec{r}_j|;R) - \frac{\pi R^2}{v} \right] \label{eq:A}.
\end{align} 
When $A=0$, the distribution is called hyperuniform and the coefficient $B$ in Eq.~(\ref{eq:s2-2}) is given by
\begin{align}
  B=\lim_{R\to\infty}  \frac{\pi R}{v} \sum_{i,j=1}^{N} n_i n_j \left[ \a(|\Vec{r}_i-\Vec{r}_j|;R) - \frac{\pi R^2}{v} \right] \label{eq:B}.
\end{align} 

 In practice, in a finite-size system, we evaluate 
\begin{align}
  A(R)\equiv\frac{\pi}{v} \sum_{i,j=1}^{N} n_i n_j \left[ \a(|\Vec{r}_i-\Vec{r}_j|;R) - \frac{\pi R^2}{v} \right] \label{eq:AR}
\end{align} 
and
\begin{align}
  \bar{B} &\equiv\frac{1}{R_{\mathrm{max}}-R_{\mathrm{min}}} \int_{R_{\mathrm{min}}}^{R_{\mathrm{max}}} B(R) dR,  \label{eq:Bbar}\\
  B(R) &\equiv \frac{\pi R}{v} \sum_{i,j=1}^{N} n_i n_j \left[ \a(|\Vec{r}_i-\Vec{r}_j|;R) - \frac{\pi R^2}{v} \right] \label{eq:BR}
\end{align}
with $R_{\mathrm{min}}<R_{\mathrm{max}}$. In the second equation, we have taken the average over $R\in[R_{\mathrm{min}},R_{\mathrm{max}}]$ because in a hyperuniform system $B(R)$ for $R>R_{\mathrm{min}}\sim 1$ typically shows a fluctuation with $R$ around $B$.

A care must be taken to implement the summation over $i$ and $j$ in Eqs.~(\ref{eq:AR}) and (\ref{eq:BR}) for a finite-size system because, for given $i$ and $j$,  $\a(|\Vec{r}_i-\Vec{r}_j|;R)$ may count a contribution from the outside of the system.
To avoid this problem, we take a sum of $i$ over sites inside a circle of radius $L_1$ and $j$ over sites inside a circle of radius $L_2=L_1+2R_{\mathrm{max}}$,

\section{Results}\label{sec:results}

We first discuss how the electron-electron interactions, $U$ and $V$, change the charge distributions in Secs.~\ref{ssec:u} and \ref{ssec:v}. We then show the resultant charge-distribution maps in real (Sec.~\ref{ssec:rmap}) and perpendicular (Sec.~\ref{ssec:perp_result}) spaces. The map in the Fourier space is presented in Appendix \ref{sec:qmap}.
In Secs.~\ref{ssec:n2} and \ref{ssec:ma}, we analyze the  charge distributions in terms of the local density variance and mulitfractality, respectively.
In Sec.~\ref{ssec:hyper_result}, we characterize the various charge distributions in terms of hyperuniformity.
Real-space patterns of the local density of states and a hyperuniformity analysis of Aubry-Andr\'e-Harper model are presented in Appendices \ref{sec:ldos} and \ref{sec:aah}, respectively.

\subsection{Effect of $U$}\label{ssec:u}
\begin{figure}[tb]
\center{
\includegraphics[width=0.48\textwidth]{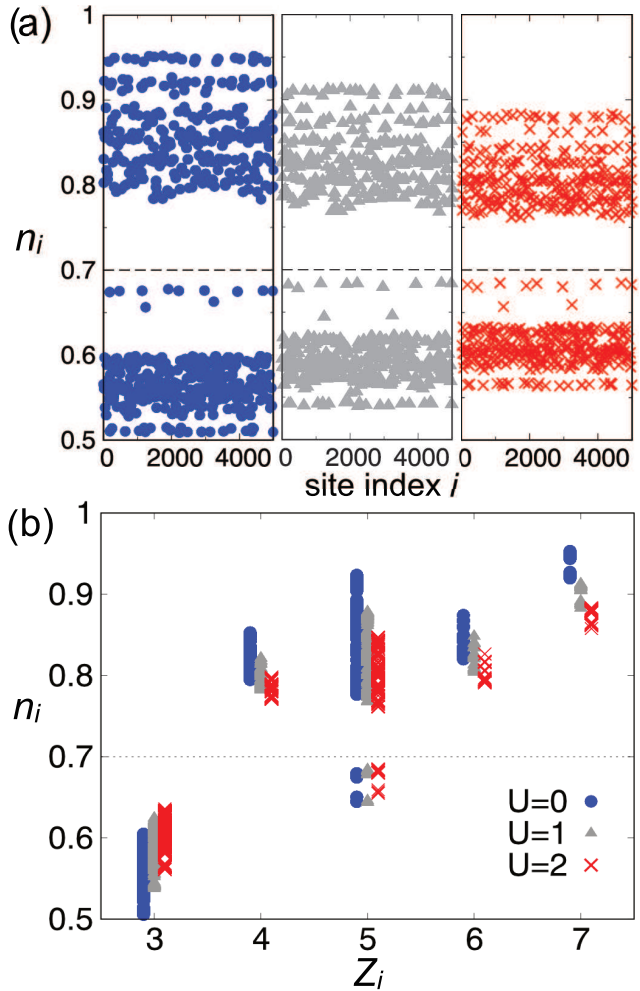}}
\caption{(a) Site-dependent electron density $n_i$ plotted against the site index $i$ for $U=0$  (left), $U=1$ (middle), and $U=2$ (right) for $V=0$ and $\nb=0.7$. The calculation was done for $N=11006$ sites, among which the central 5000 sites are plotted to avoid a boundary effect. The horizontal dashed line denotes $\nb$. The sites are indexed from 1 to $N$, according to the following rules: (i) The index of the central site is 0. (ii) The smaller index is given to a site closer to the center in the Euclidean distance $|\Vec{r}|$ [see Fig.~\ref{fig:penrose}(a)]. (iii) For the sites with the same distance from the center, a smaller index is given to a site with a smaller anticlockwise angle measured from the right direction. (b) The electron density $n_i$ plotted against the coordination number $Z_i$. The data points are slightly shifted in the horizontal direction for the sake of visibility. Only the inner sites satisfying $|\Vec{r}|<45$ are used.}
\label{fig:udep}
\end{figure}

Already in the non-interacting limit $U=V=0$, the electron density distributes nonuniformly [Fig.~\ref{fig:udep}(a), left panel] due to the effect of $t$. 
For $\nb<1$, a site with a larger coordination number $Z_i$ tends to have more electrons [Fig.~\ref{fig:udep}(b)].
This is because of a larger benefit of the kinetic energy at the sites with a larger $Z_i$.

As $U$ increases with $V=0$ fixed, this charge modulation is suppressed [Fig.~\ref{fig:udep}(a)]. This is because $U>0$ prefers a more uniform distribution to reduce the onsite-interaction energy.
Note that for $V=0$, $\nb>1$ and $\nb<1$ are related through the electron-hole transformation, so that a site with a larger $Z_i$ tends to have a smaller electron density for $\nb>1$.

\subsection{Effect of $V$}\label{ssec:v}
\begin{figure}[tb]
\center{
\includegraphics[width=0.48\textwidth]{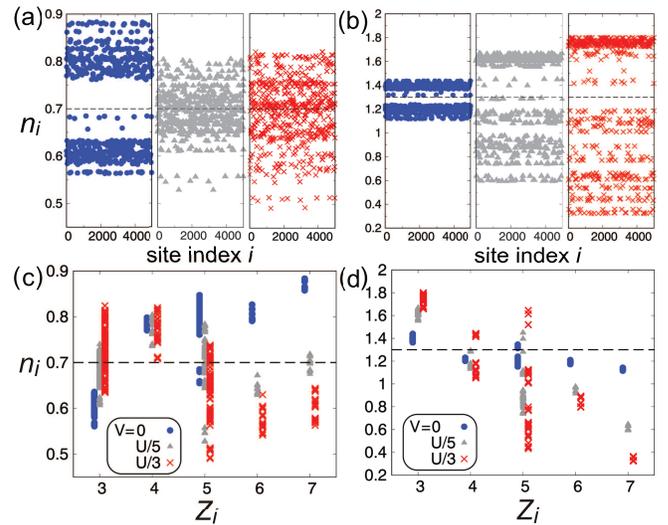}}
\caption{ (a) Site-dependent electron density $n_i$ plotted against the site index $i$ for $V=0$  (left), $V=U/5$ (middle), and $V=U/3$ (right) for $U=2$ and $\nb=0.7$. (b) The same for $\nb=1.3$. (c) and (d) plot $n_i$ against $Z_i$. The numbers of the sites used for the calculation and for the plot, as well as the site-indexing rule, are the same as those used in Fig.~\ref{fig:udep}.}
\label{fig:vdep}
\end{figure}

In the presence of the intersite interaction $V$, the situation changes.
As shown in Fig.~\ref{fig:vdep}(a) for $\nb=0.7$, while $V=U/5$ suppresses the charge modulation,  $V=U/3$ induces a prominent charge modulation. 
For $\nb=1.3$ [Fig.~\ref{fig:vdep}(b)], on the other hand, $V$ monotonically enhances the modulation.
Figures \ref{fig:vdep}(c) and \ref{fig:vdep}(d) show that, in these $V$-induced charge distributions, electrons tend to populate the sites with smaller $Z_i$ to suppress the energy increase due to the $V$ term in Eq.~(\ref{eq:mf}).
For $\nb<1$, this population tendency is opposite to that due to $t$ seen for $V=0$.
This is why the modulation is suppressed for $V=U/5$ and $\nb=0.7$; there is a competition of two opposite tendencies. For $V=U/3$, the effect of $V$ prevails, inducing a different type of modulation.
For $\nb>1$, on the other hand, the population tendency due to $V$ matches that of $t$, so that they work cooperatively to result in the large charge modulation.

The effect of $V$ discussed so far can be interpreted as the contribution from the Hartree term in Eq.~(\ref{eq:mf}). 
In fact, in Appendix \ref{sec:fock}, we will see that the Fock term does not play a significant role in these charge distributions. 

\subsection{Real-space map} \label{ssec:rmap}
\begin{figure*}[tb]
\center{
\includegraphics[width=0.9\textwidth]{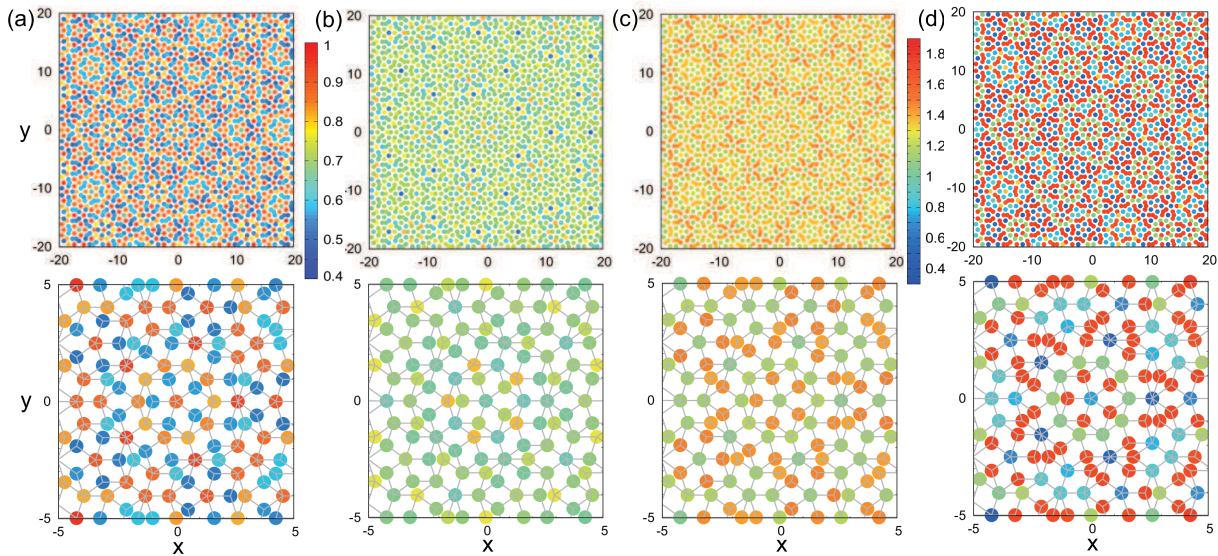}}
\caption{Real-space map of the electron density for (a) $\nb=0.7$ and $U=V=0$, (b) $\nb=0.7$ and $U=4V=2$, (c) $\nb=1.3$ and $U=V=0$, and (d) $\nb=1.3$ and $U=4V=2$. Panels (a) and (b) [(c) and (d)] share the color scale. The calculations were done for a cluster with $N=11006$ sites, among which the central area satisfying $|x|, |y|<20 (5)$ is plotted in the top (bottom) panels.}
\label{fig:rmap}
\end{figure*}

Figure \ref{fig:rmap} shows a real-space map of $n_i$. 
It presents interesting spatial patterns already for $U=V=0$ [panels (a) and (c)] due to the effect of $t$.
A similar but weaker modulation occurs for $U>0$ and $V=0$ (not shown), in accordance with the behavior seen in Fig.~\ref{fig:udep}.
The introduction of $V$, on the other hand, changes the spatial patterns drastically [panels (b) and (d)]. 
This is because the population at each site changes differently according to its local geometry,  as we have seen above.
For instance, in the lower panels of Figs.~\ref{fig:rmap}(a) and \ref{fig:rmap}(b), the site at the center loses a population (red $\to$ light blue) while the sites surrounding it gain a population (blue $\to$ light green).


\begin{figure}[tb]
\center{
\includegraphics[width=0.48\textwidth]{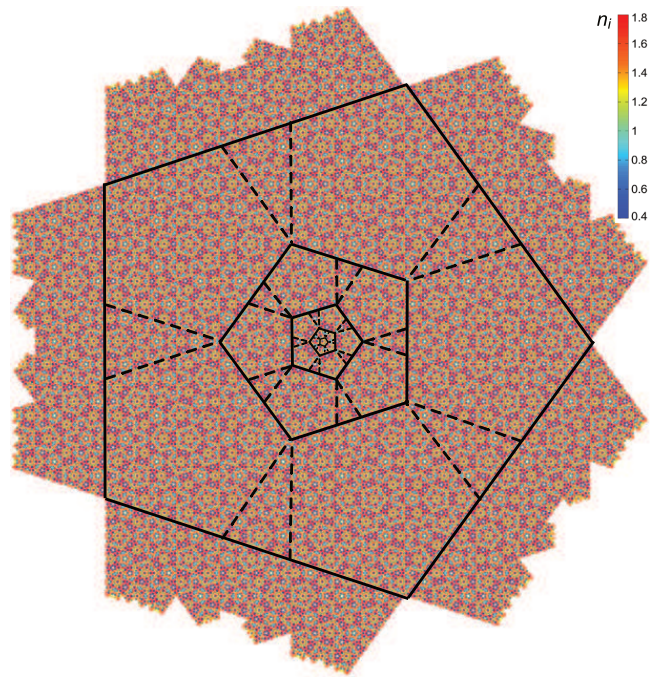}}
\caption{Self-similar charge distribution on the Penrose tiling. The calculation was done for $N=75806$, $\bar{n}=1.3$ and $U=4V=2$. The black (solid and dashed) lines are a guide for the eye to see a self-similarity.
} \label{fig:frac}
\end{figure}

In a more global view, these real-space maps exhibit a self-similarity, as seen in Fig.~\ref{fig:frac}:
We can see a pentagonal area (denoted by black lines) successively enlarged by $\tau^2\simeq 2.618$ times, where  $\tau\equiv \frac{1+\sqrt{5}}{2}$ is the golden ratio. 
This self-similar structure reflects that of the underlying lattice and is analogous to that seen in  the wave functions in the noninteracting system \cite{sutherland87,tokihiro88}. 
Nevertheless, unlike the wave functions, the charge distribution does not show the multifractality: Although we have computed the multifractal dimension for $n_i$, it always approaches a trivial number (i.e., equal to the spatial dimension) in the thermodynamic limit (see Sec.~\ref{ssec:ma}).
The difference would be attributed to the different variable ranges:
Namely, $n_i$ is limited to $[0,2]$ at each site, so that its fluctuation cannot be so large compared to its mean value while such a limitation is absent for the wave-function amplitude.
We also note that by summing the occupied states, the multifractal nature of the eigenfunction
is smeared out.

\subsection{Perpendicular-space profile}\label{ssec:perp_result}

\begin{figure*}[tb]
\center{
\includegraphics[width=0.9\textwidth]{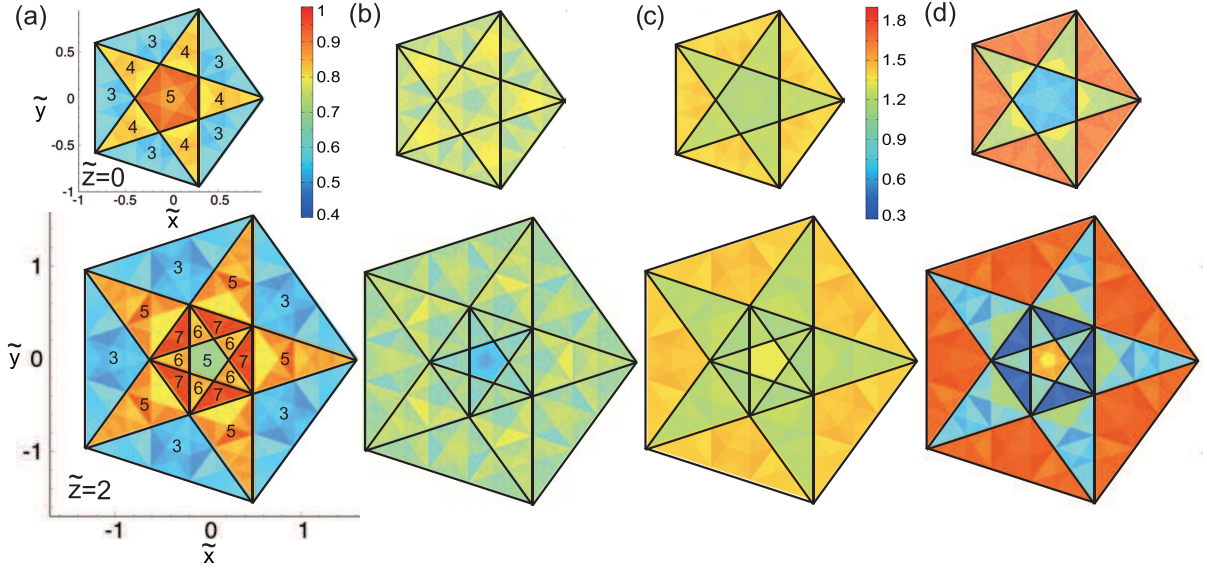}}
\caption{Intensity map of the electron density in the perpendicular space for (a) $\nb=0.7$ and $U=V=0$, (b) $\nb=0.7$ and $U=4V=2$, (c) $\nb=1.3$ and $U=V=0$ and (d) $\nb=1.3$ and $U=4V=2$. The calculations were done for $N=520851$, among which the sites satisfying $|\Vec{r}|<320$ are used to avoid the effect of the boundary.
(a) and (b) [(c) and (d)] share the same color scale. 
We have overlapped the black lines dividing the perpendicular space into sections. In panel (a), the number in each section denotes the coordination numbers $Z_i$ of the sites in the physical space.}
\label{fig:perp}
\end{figure*}

The various charge distribution patterns presented above would be attributed to the different effects of $t$, $U$, and $V$ at sites with different local geometries, as indicated by the results in Secs.~\ref{ssec:u} and \ref{ssec:v}. 
Such a dependence on the local geometry can be more explicitly seen in the perpendicular space (Sec.~\ref{ssec:perp}), where the sites with a similar local geometry are arranged in the same area. 

Figure \ref{fig:perp} presents the perpendicular-space profile calculated for $N=520851$. 
Here, we have shown the results only for $\tilde{z}=0$ and $2$ since $\tilde{z}=1$ and $3$ are equivalent to these:
Although it is an intriguing possibility that the effect of $V$ may break the sublattice symmetry, we have not found such a symmetry breaking.
It is known that each section [separated by black lines in panel (a)] corresponds to the sites in a specific nearest-neighbor geometry in the physical space \cite{bruijn81-1,bruijn81-2}. 
Fine structures within each section correspond to the geometry beyond the nearest neighbors.
In panel (a), we have denoted the coordination number $Z_i$ of each section.
 
For $\nb=0.7$ and $U=V=0$ [Fig.~\ref{fig:perp}(a)], the sites with $Z_i=3$ ($Z_i=7$) has a relatively small (large) $n_i$, in consistency with Fig.~\ref{fig:udep}(b).
Although some fine structures are discernible in each section, the overall trend is that the sections with $Z_i=3$ have a smaller density while those with $Z_i \geq 4$ have a larger density.
The opposite trend is seen for $\nb=1.3$ and $U=V=0$ [Fig.~\ref{fig:perp}(c)].

\begin{figure}[tb]
\center{
\includegraphics[width=0.4\textwidth]{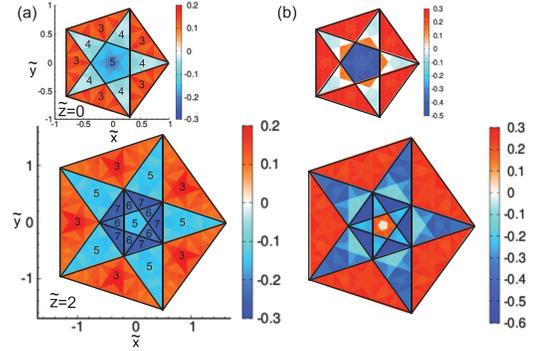}}
\caption{Change of the electron density $\Delta n_i\equiv n_i^{(U=4V=2)}-n_i^{(U=V=0)}$ plotted in the perpendicular space for (a) $\nb=0.7$ and (b) $\nb=1.3$. }
\label{fig:perp_diff}
\end{figure}

The interaction effect considerably changes the profile. 
The change is more clearly seen in Fig.~\ref{fig:perp_diff}, which plots the density difference $\Delta n_i$ between $U=4V=2$ and $U=V=0$.
For $\nb=0.7$, the sites with $Z_i \geq 4$ lose the population while the $Z_i=3$ sites gain the population.
For $\nb=1.3$, the $Z_i=3$ sites and a part of the $Z_i=4$ and $Z_i=5$ sites gain a population while the other sites lose it. 
Here, the $Z_i=5$ section at the center of the $\tilde{z}=2$ plane corresponds to the sites of the S5 vertex (see Fig.~\ref{fig:z5} below).

\begin{figure}[tb]
\center{
\includegraphics[width=0.48\textwidth]{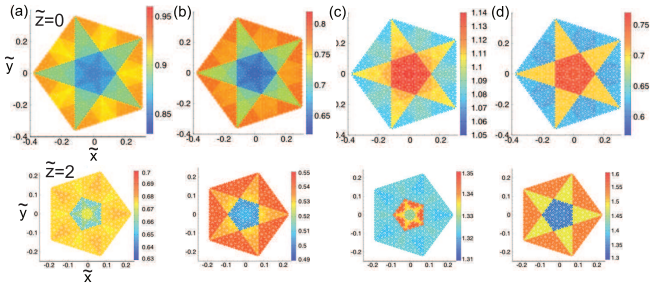}}
\caption{Enlarged views of the $Z_i=5$ sections at the center of the upper and lower panels in Fig.~\ref{fig:perp}.}
\label{fig:perp2}
\end{figure}

\begin{figure}[tb]
\center{
\includegraphics[width=0.4\textwidth]{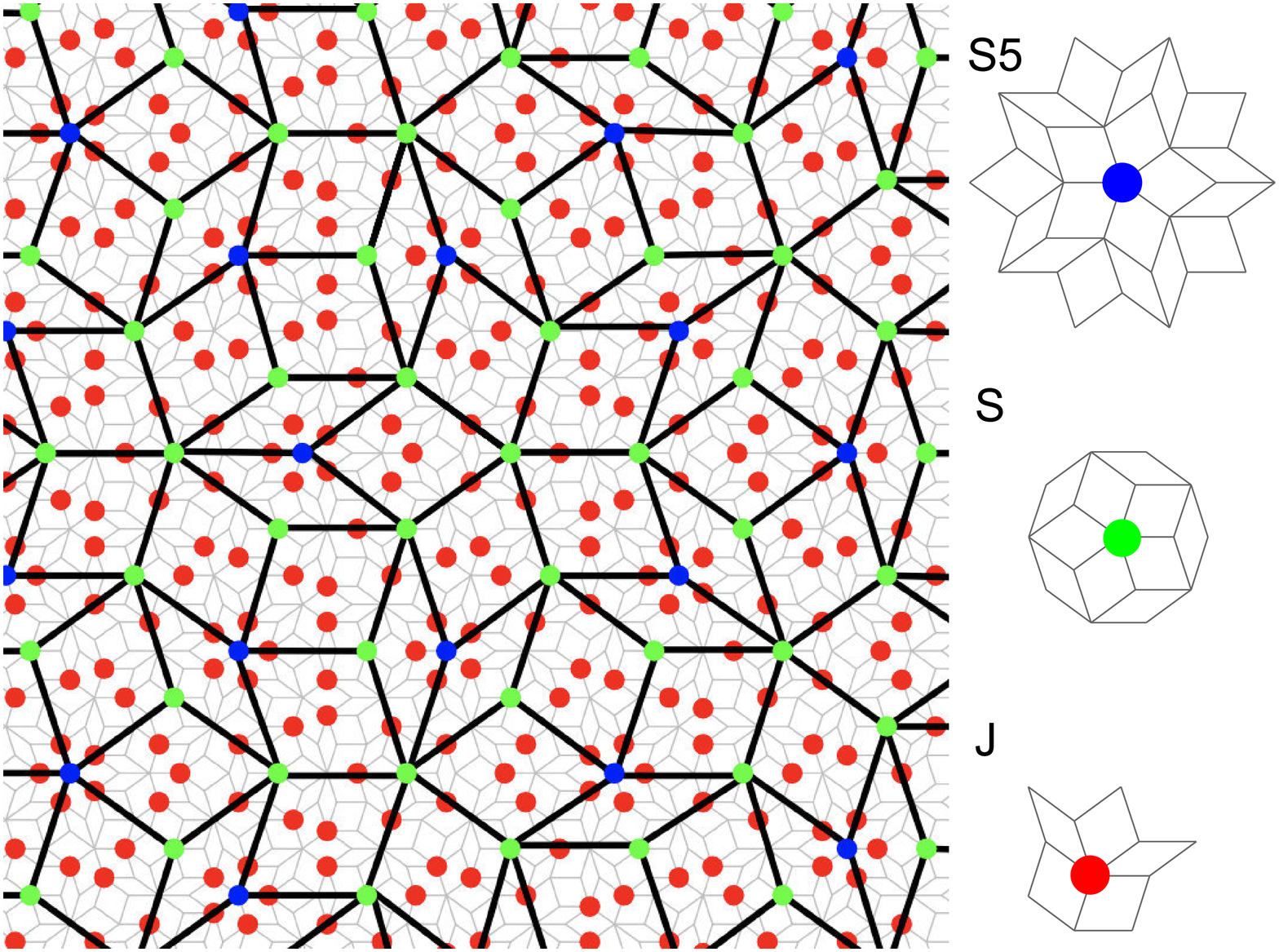}}
\caption{Self-similarlity of the Penrose tiling. Blue, light green, and red circles denote the $Z_i=5$ sites with the local geometries shown below (S5, S, and J vertices in the nomenclature by de Brujin \cite{bruijn81-1,bruijn81-2}). The thin gray lines represent the original lattice while the bold black lines connects the S and S5 vertices.}
\label{fig:z5}
\end{figure}

Interestingly, the perpendicular-space profile also shows a structure reflecting the self-similarity of the Penrose lattice. Looking into the $Z_i=5$ sections at the center of the $\tilde{z}=0,2$ planes, we find a fine structure that resembles the original perpendicular-space profile:
As Fig.~\ref{fig:perp2} shows,  there appear a star and a pentagon {\it inside} the $Z_i=5$ sections although the modulation is rather weak.
Such a self-similar structure in the perpendicular space has been seen in the magnetization profile in the antiferromagnetic state on the Ammann-Beenker tiling \cite{wessel05,jagannathan12,koga20}, where the symmetric $Z_i=8$ vertices constitute a "superlattice" of larger tiles in a self-similar manner.

Here, on the Penrose tiling, the symmetric $Z_i=5$ vertices (S and S5 vertices) constitute the "superlattice" as shown in Fig.~\ref{fig:z5}. 
Each large rhomboidal tile consists of one S5 vertex and three S vertices of the original tiling.
The length scale of this new tile is $\tau^3$ times larger than that of the original tiles. 
Because the effective hopping, as well as the effective intersite interaction, between the neighboring sites on this superlattice will be nonuniform, the charge-distribution pattern does not show a complete self-similarity. In addition, differently from the Ammann-Beenker tiling, where the the symmetric vertices have the largest coordination number $Z_i=8$, the symmetric vertices in the Penrose tiling have an intermediate coordination number $Z_i=5$, likely making the hierarchical structure more ambiguous. 
Nevertheless, it is interesting that we can still discern a fine structure reflecting the self-similarity of the underlying lattice.

\subsection{Local density variance}\label{ssec:n2}

\begin{figure}[tb]
\center{
\includegraphics[width=0.48\textwidth]{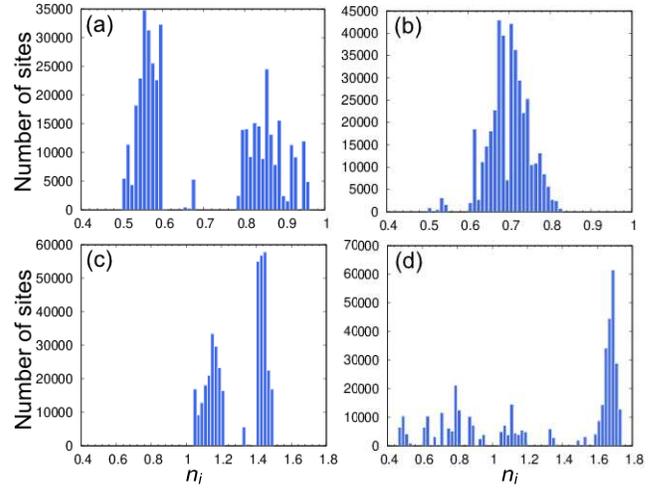}}
\caption{ Distribution of the electron density for (a) $\nb=0.7$ and $U=V=0$, (b) $\nb=0.7$ and $U=4V=2$, (c) $\nb=1.3$ and $U=V=0$, and (d) $\nb=1.3$ and $U=4V=2$, obtained for a cluster of 520851 sites, among which the sites satisfying $|\Vec{r}| < 320$ are used to avoid the boundary sites. The bin width is 0.01 (0.02) for (a) and (b) [(c) and (d)]. }
\label{fig:dist}
\end{figure}

Figure \ref{fig:dist} shows how the histgram of $n_i$ changes with the interactions. 
For $\nb=0.7$ and $U=V=0$ [panel (a)], we see that $n_i$'s are classified into roughly two groups, around 0.5-0.6 and around 0.8-0.9, with a gap around 0.7. 
The interaction $U=4V=2$ moves most of the points to the range 0.6-0.8, suppressing the spread [panel (b)]. On the other hand, for $\nb=1.3$, $n_i$ gets more distributed as the interaction is introduced. According to Fig.~\ref{fig:vdep}(d), the large peak around $n_i=1.7$ consists of the sites with $Z_i=3$ while the small peak around $n_i=0.4$ consists of the $Z_i=7$ sites. 
The former large peak reflects the fact that more than half of the lattice points have $Z_i=3$ [Fig.~\ref{fig:penrose}(c)].

\begin{figure}[tb]
\center{
\includegraphics[width=0.48\textwidth]{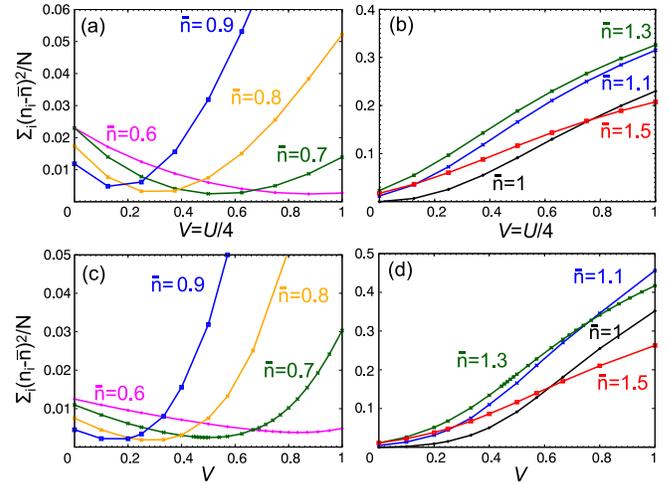}}
\caption{Local density variance plotted against (a), (b) $U=4V$ and (c), (d) $V$ with $U=2$ fixed. The calculations were done for 11006 sites, among which we have used only the sites satisfying $|\Vec{r}|<45$ for the plot.}
\label{fig:n2}
\end{figure}

In order to quantify this spread, we define the local density variance by
\begin{align}
\frac{1}{N}\sum_i (n_i-\nb)^2. \label{eq:n2}
\end{align}
Figure \ref{fig:n2} plots this quantity against $U=4V$ (upper panels) and against $V$ with $U=2$ fixed (lower panels).
First of all, we notice that the variance is one order of magnitude larger for $\nb>1$ than for $\nb<1$, breaking the electron-hole symmetry.
This is simply because the $V$ term in the Hamiltonian (\ref{eq:hubbard}) is quadratic in density.

In more detail, for $\nb<1$, the curves are convex down and nonmonotonic, reflecting the competition between the effects of $t$ and $V$.  The minimum, where the system is closest to the uniform distribution, shifts to a smaller value of $V$ as $\nb$ increases. This is because the effect of $V$ gets stronger as $\nb$ increases. 
The minimum reaches $V=0$ at $\nb=1$.
For $\nb\ge 1$, the curves monotonically increase with $V$ and show an inflection point. Interestingly, it is not always monotonic as a function of $\nb$: In Fig.~\ref{fig:n2}(b), the variance at $\nb=1.3$ is the largest. This will be ascribed to another competition between the increasing effect of the $V$ term and the decrease of the vacant sites as $\nb$ increases.

Thus, the local density variance captures several interesting features of the distribution while it does not reflect the spatially ordered feature of the charge distribution, as it is a spatially averaged local fluctuation and does not directly reflect the underlying lattice structure:  
 The same value of the variance may be realized by a random distribution, too: For example, if we consider a random but statistically homogeneous distribution of $n_i$, we can tune the local density variance with keeping the average electron density.

\subsection{Multifractal analysis}\label{ssec:ma}
\begin{figure}[tb]
\center{
\includegraphics[width=0.48\textwidth]{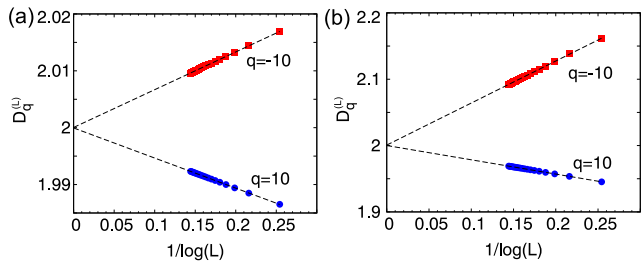}}
\caption{$L$ dependence of $D_q^{(L)}$ at $q=\pm10$, calculated at $\nb=1.3$ for (a) $U=V=0$ and (b) $U=4V=2$. Calculation was done for a cluster with 1~364~431 sites, among which only the sites satisfying $|\Vec{r}|<520$ are used. Dashed lines show a  linear fitting to the data.}
\label{fig:Dq}
\end{figure}

We have calculated the multifractal dimension for the obtained charge distributions. However, we do not find any meaningful deviation of $D_q$ from the spatial dimension $d=2$ in the thermodynamic limit.

Figure \ref{fig:Dq} plots $D_q^{(L)}$ against $1/\log(L)$, calculated for two different parameter sets. 
Because $D_q$ is generally a non-increasing function with respect to $q$, it is bounded by the values at $q=\pm\infty$. 
In addition, $D_q$ is nearly flat for a large $|q|$, where a site with maximum or minimum $n_i$ dominates its value.
In the figure, we therefore plot $D_q^{(L)}$ at $q=\pm10$, which are considered to be sufficiently large.

In both cases, $D_q^{(L)}$ approaches $d=2$ in the large $L$ limit with a logarithmically slow convergence against $L$. We have obtained similar results for other values of $q$ and model parameters. We therefore conclude that the charge distribution on the Penrose lattice is {\it not} multifractal.

This result may be understood as follows.
As we have seen in Sec.~\ref{ssec:perp_result}, $n_i$ is mostly determined by a short-range geometry. Therefore, if we denote by $r_g$ the ratio of the sites in each short-range geometry $g$, the sums in Eq.~(\ref{eq:norm}) can be approximated as
\begin{align}
    \sum_i n_i\simeq N\sum_g r_g n_g,\\
    \sum_i n_i^q\simeq N\sum_g r_g n_g^q,
\end{align}
where $n_g$ denotes $n_i$ of the site $i$ in the geometry $g$.
Then, Eq.~(\ref{eq:norm}) is approximated as 
\begin{align}
    ||n||_q^{(L)}\simeq N^{1-q}\frac{\sum_g r_g n_g^q}{(\sum_g r_g n_g)^q}.
\end{align}
Because $r_g$ is independent of $N\propto L^2$ and the sum $\sum_g$ is over a {\it finite} number of  short-range geometries, we obtain
\begin{align}
    \ln||n||_q^{(L)}\simeq 2(1-q)\ln L + {\rm const.}
\end{align}
Putting this into Eq.~(\ref{eq:dql}), we obtain $D_q^{L}$ logarithmically approaching to $d=2$.

\subsection{Hyperuniformity}\label{ssec:hyper_result}
\begin{figure}[tb]
\center{
\includegraphics[width=0.48\textwidth]{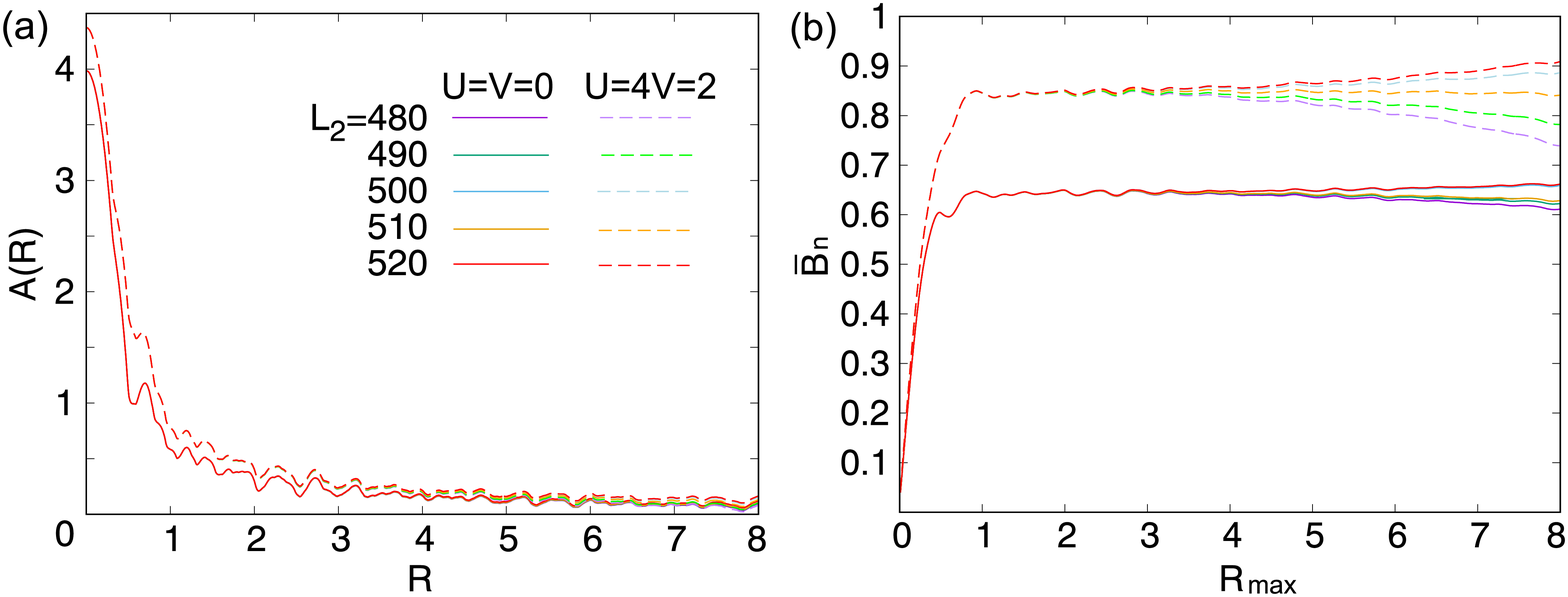}}
\caption{(a) $A(R)$ plotted against $R$. (b) $\bar{b}_{\mathrm{n}}(R_\mathrm{max})$ plotted against $R_\mathrm{max}$. Calculation was done for $\nb=1.3$ with different values of $L_2$ from 480 to 520.
Solid (dashed) curves show the results for $U=V=0$ ($U=4V=2$). Note that the curves of all $L_2$ values are overlapping in panel (a).}
\label{fig:hu_r}
\end{figure}

Here, we apply the idea of the hyperuniformity (Sec.\ref{ssec:hyper}) \cite{torquato03,torquato16,torquato18}, which captures the underlying lattice structure, distinguishing it from a random distribution, and would give a measure of the regularity of the density distribution.
This idea will be particularly useful when a distribution is neither characterized by symmetry breaking nor multifractality, as is the present case.

To see if a distribution is hyperuniform or not, we examine whether $A(R)$ [Eq.~(\ref{eq:AR})] disappears at a large $R$.
When it disappears, the hyperuniformity order metric $\bar{B}$ [Eq.~(\ref{eq:Bbar})] gives a measure of the regularity.
As is clear from Eqs.~(\ref{eq:AR}) and (\ref{eq:Bbar}), these quantities reflect a nonlocal feature of the distribution beyond the local fluctuation taken into account by Eq.~(\ref{eq:n2}). 

Figure \ref{fig:hu_r} shows the $R$ ($R_\mathrm{max}$) dependence of $A(R)$ ($\bar{B}$) for $U=V=0$ and $U=4V=2$ at $\nb=1.3$. 
Here, we have performed the Hartree-Fock calculation for a 1~364~431-site cluster and used a region inside a circle of the radius $L_2$ ($480\leq L_2 \leq 520$) around the center for calculating $A(R)$ and $\bar{B}$.
For $\bar{B}$, we have plotted a normalized quantity, $\bar{B}_{\mathrm{n}}\equiv \bar{B}/(\phi^{1/2} \nb^2)$
with $\phi\equiv\frac{\pi N}{4v} $\footnote{This $\phi$ factor is introduced to make the definition consistent with Ref.~\onlinecite{torquato16}, where the order metrics of various lattices are compared} to eliminate the trivial factor scaling with $\nb^2$.

The decrease of $A(R)$ to zero with increasing $R$ shows that the density distribution is hyperuniform for both $U=V=0$ and $U=4V=2$.
$\bar{B}_{\mathrm{n}}$ is nearly flat with a small fluctuation for $1\lesssim R_\mathrm{max}\lesssim 5$.
However, for $R_\mathrm{max} \gtrsim 5$, $\bar{B}_{\mathrm{n}}$ shows an upward or downward shift depending on $L_2$, and this shift becomes large as $R_\mathrm{max}$ increases. 
This will be attributed to the finite system size.

\begin{figure}[tb]
\center{
\includegraphics[width=0.48\textwidth]{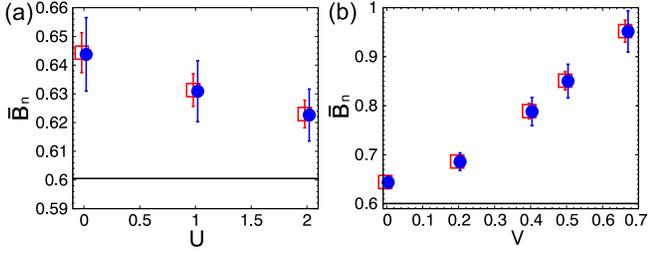}}
\caption{Interaction-parameter dependence of $\bar{B}_\mathrm{n}$, calculated for $\nb=1.3$. (a) $U$ dependence for $V=0$. (b) $V$ dependence for $U=2$. Two averages over $470<L_2\leq520$ (red square) and $270<L_2 \leq320$ (blue circle) are also compared, where we have slightly shifted the data points to the horizontal direction just for the visibility. The horizontal black line shows the value  $\bar{B}_{\mathrm{n}p}$ for a point set (i.e., $n_i\equiv 1$) }
\label{fig:hu_n13}
\end{figure}

We therefore take $R_{\mathrm{max}}=5$ to evaluate $\bar{B}_{\mathrm{n}}$, where we take the average over 50 samples of $L_2$ ($=471, 472, ... , 520$ for $N=1364431$ and $=271, 472, ... , 320$ for $N=520851$) and evaluate the error bar.
The results are presented in Fig.~\ref{fig:hu_n13} for $\nb=1.3$.
We see that the two averages of different ranges of $L_2$ give almost the same result except that the error bar is smaller for the larger $L_2$ range.
Note that the order metric of the point set (i.e., $n_i\equiv 1$) is about $\bar{B}_{\mathrm{n}p}\equiv 0.60052$ \cite{zachary09,torquato18}, which is well reproduced by the present calculation and shown by the horizontal line.

For $U=V=0$ [see Fig.~\ref{fig:hu_n13}(a)], $\bar{B}_{\mathrm{n}}$ is about 0.644, which is substantially larger than $\bar{B}_{\mathrm{n}p}$ due to the modulation induced by $t$.
As $U$ increases, $\bar{B}_{\mathrm{n}}$ decreases monotonically in accord with the intuition that the density fluctuation is suppressed:
It will approach $\bar{B}_{\mathrm{n}p}$ in the large $U$ limit.
When $V$ is introduced with $U=2$ fixed [Fig.~\ref{fig:hu_n13}(b)], $\bar{B}_{\mathrm{n}}$ increases significantly.
Thus, the $V$-induced distribution pattern seen in Fig.~\ref{fig:rmap}(d) is characterized by the order metric much larger than that of the $t$-induced patterns at $V=0$ [Fig.~\ref{fig:rmap}(c)].

\begin{figure}[tb]
\center{
\includegraphics[width=0.48\textwidth]{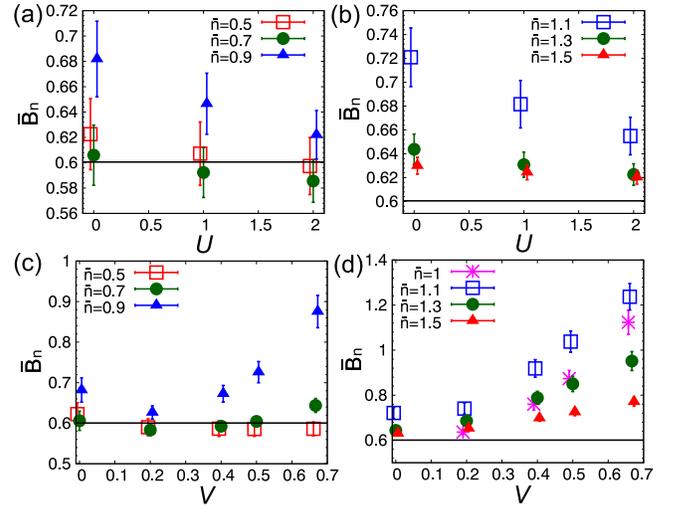}}
\caption{Interaction-parameter dependence of $\bar{B}_{\mathrm{n}}$, calculated for various $\nb$. (a) [(b)] $U$ dependence for $\nb<1$ ($\nb>1$) at $V=0$, (c) [(d)] $V$ dependence for $\nb<1$ ($\nb\geq 1$) at $U=2$.  $\bar{B}$ is averaged over $270<L_2\leq320$.  Each data point is slightly shifted to the horizontal direction just for visibility.}
\label{fig:hu_n}
\end{figure}

In Fig.~\ref{fig:hu_n}, we explore the interaction effects on the hyperuniformity for various electron fillings.
For $\nb\leq0.7$,  $\bar{B}_{\mathrm{n}}$ does not change much with $t$, $U$, and $V$: The change is mostly within the error bar.
For $\nb\geq0.9$, we see that $t$ substantially enhances $\bar{B}_{\mathrm{n}}$ while $U$ suppresses it.
The effect of $V$ is nonmonotonic for $\nb=0.9$ while a pronounced and monotonic increase of $\bar{B}_{\mathrm{n}}$ is seen for $\nb\geq 1$.
Interestingly, as a function of $\nb$, $\bar{B}_{\mathrm{n}}$ takes a maximum around $\nb=1$ for a large $V$.
This will be because $n_i$ can fluctuate most strongly around $\nb=1$ in the present mean-field approach.
These behaviors are consistent with the results obtained in the previous sections, as well as with our intuition, demonstrating that the hyperuniformity order metric is a reasonable measure to quantify the inhomogeneous density distributions in quasiperiodic systems.

\section{Discussion}\label{sec:discuss}

While it has been known that the distribution of the lattice points in quasiperiodic lattices is hyperuniform \cite{zachary09,oguz17,torquato18}, the present study revealed that the density distributions on them are also hyperuniform. 
Since hyperuniformity means that the density variance is contributed only from the surface area, the obtained results mean that the density contained in the bulk region does not fluctuate, unlike a random distribution. 
Such a property trivially holds on periodic lattices while it is not so trivial for quasiperiodic lattices to have this property.
This should be ascribed to the regularity of the quasiperiodic lattices.
In the high-dimensional construction of the quasiperiodic lattices, when we consider a sufficiently large system (window) in the physical space, the corresponding coordinates in the perpendicular space uniformly distribute \cite{senechal95book}.
Since each position in the perpendicular space corresponds to a specific local geometry in the physical space, the uniform distribution in the former space means that the variety and the ratio of the local geometry in the latter space do not depend on the position of the window.
Combining this with our observation that $n_i$ is mostly determined by the local geometry (Sec.~\ref{ssec:perp_result}), we can understand that the density inside the window does not have a bulk contribution.

Studying the Aubry-Andr\'e-Harper model \cite{aubry80,harper55} as another type of the quasiperiodic system, we have confirmed a similar behavior, i.e., hyperuniform density distribution. Note that the wave function in this model is known to be extended below (localized above) a critical strength of the potential. Applying the method described in Sec.~\ref{ssec:hyper} to the amplitude of these wave functions, we have confirmed that the extended wave function is hyperuniform while the critical and localized wave functions are not. We also find that the charge distribution changes from the Class-I hyperuniform to Class-II hyperuniform \cite{torquato18} at the self-dual point. For more details, see Appendix \ref{sec:aah}. 

Although the electron density on any periodic lattice also shows a hyperuniformity, it is usually fixed at the value of the point set unless the charge order, characterized by the broken translational symmetry, occurs.
On the other hand, in quasiperiodic systems, the distribution changes with the change of various parameters, such as the chemical potential, hopping integrals, and the strength of the electron-electron interactions, without any signal of the change of the symmetry: There is no translational symmetry in the first place. 
Then, the different distributions give different order metrics of hyperuniformity.
Thus, the electron states in quasiperiodic systems are an interesting playground of hyperuniformity, which is controllable by the above-mentioned parameters.

It will also be interesting to think about different local isomorphism (LI) classes of the generalized Penrose tiling. In Refs.~\onlinecite{lin17} and \onlinecite{lin18}, it was clarified that the point patterns of different LI classes result in different hyperuniformity order metrics and different localization properties of light on corresponding photonic quasicrystals. Therefore, different LI classes may have different consequences on the hyperuniformity of the charge distributions and their physical properties, too. This is an intriguing future research direction.

Another intriguing future issue will be the effect of disorders on the hyperuniformity of the charge distribution. The effect of imperfections on various point patterns was studied in Ref.~\onlinecite{kim18}, which revealed that imperfections can destroy the hyperuniformity or alter its class. Their effects on hyperuniform electronic states are therefore worth investigation.

\section{summary}\label{sec:summary}

We have studied the effect of electron-electron interactions on the metallic state of quasicrystals.
Introducing the extended Hubbard model on the Penrose tiling, we determined the charge distribution self-consistently within the Hartree-Fock approximation.
The charge distribution shows a self-similarity, reflecting the underlying quasiperiodic lattice structure.
We have found that the onsite repulsion $U$ suppresses this charge inhomogeneity while the intersite interaction $V$ can significantly enhance the inhomogeneity, leading to distinct spatial patterns.
The strong effect of $V$ is attributed to the variation of the local geometries around each site.  
It also shows a strong electron-hole asymmetry, resulting in a much larger effect on the electron-rich side than the hole-rich side. 
This is substantiated by a systematic study of the local density variance quantifying the inhomogeneity: 
It shows a nonmonotonic dependence on $V$ on the hole-rich side and a rapid monotonic increase on the electron-rich side.

Despite the seemingly self-similar structure, these charge distribution patterns do not show the multifractality.
We have then applied the idea of hyperuniformity \cite{torquato03,torquato16,torquato18} to characterize the global feature of the distributions, which is not taken into account by the local density variance.
We have shown that the density distribution is indeed hyperuniform and that its order metric systematically changes with the model parameters, in a way consistent with our intuition about the distribution. In particular, the intersite interaction substantially increases the order metric.
These results show that the hyperuniformity order metric works as a measure to quantify these nonmultifractal density distributions on quasiperiodic lattices.

Aside from the electron density, interesting spatial patterns have been reported in preceding theoretical studies for the magnetic moment \cite{wessel03,vedmedenko04,jagannathan04,wessel05,jagannathan07,szallas09,jagannathan12,thiem15,koga17,koga20,hauck21,watanabe21} of  antiferromagnets,  superconducting order parameter \cite{sakai17,araujo19,sakai19,cao20,zhang20,nagai20,takemori20,nagai21}, and the order parameter of the excitonic insulator \cite{inayoshi20}.
The concept of hyperuniformity may also be useful to characterize these distributions.

\begin{acknowledgments}
We thank Masatoshi Imada, Anuradha Jagannathan, Akihisa Koga, Masaki Tezuka, Paul J. Steinhardt, and Salvatore Torquato for valuable comments and discussions.
This work was supported by JSPS KAKENHI Grant No. JP16H06345, JP19H00658, JP19H05825 and JP20H05279.
\end{acknowledgments}

\bibliography{ref}

\appendix \section{Effect of the Fock term}\label{sec:fock}
\begin{figure}[tb]
\center{
\includegraphics[width=0.48\textwidth]{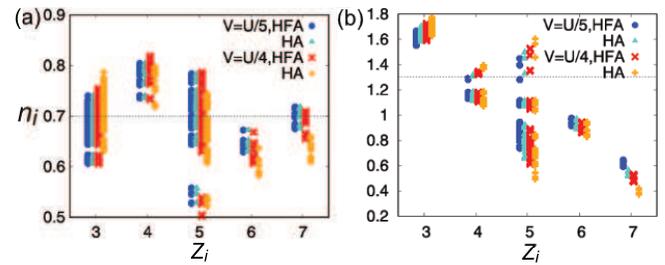}}
\caption{(a), (b) Comparison of $n_i$ obtained with the Hartree-Fock approximation (HFA) and Hartree approximation (HA) for $\nb=0.7$ and $1.3$, respectively.
The calculations were done for $U=2$ and $N=11006$, among which we have used only the inner sites satisfying $|\Vec{r}|<45$ for the plot.}
\label{fig:fock}
\end{figure}

In Fig.~\ref{fig:fock}, we can see the effect of the Fock term in Eq.~(\ref{eq:mf}), by comparing the results obtained with the Hartree (HA) and Hartree-Fock approximations (HFA). 
Overall, the results of the HFA are similar to that of the HA, indicating that the main effect of $V$ comes from the Hartree term for the parameters we studied. 
The effect of the Fock term, however, can be seen as a suppression of $|n_i-\nb|$ at $Z_i=6, 7$ for $\nb=0.7$ and $V=U/4$, and as a weak but overall suppression of the inhomogeneity for $\nb=1.3$.

\section{Momentum-space map} \label{sec:qmap}
\begin{figure}[tb]
\center{
\includegraphics[width=0.48\textwidth]{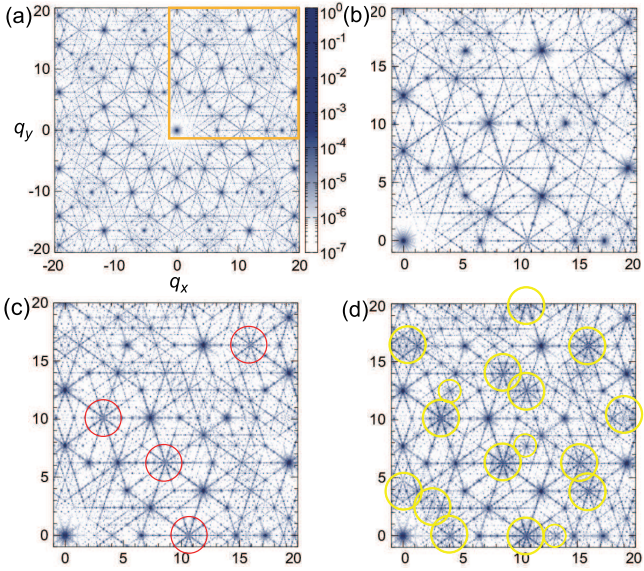}}
\caption{Momentum-space maps of the electron density for $\nb=1.3$. (a) Crystal structure factor, calculated for $U=V=0$ and for a uniform density $n_i\equiv\nb$. (b) Enlarged view of panel (a) for $x,y \in [-1,20]$ [denoted by an orange square in panel (a)]. (c) ${n_\qq}^2$ for $U=V=0$. Red circles point prominent changes from panel (b). (d) ${n_\qq}^2$ for $U=4V=2$.  Yellow circles point prominent changes from panel (c). The calculation was done for $N=520851$.
}
\label{fig:qmap}
\end{figure}

In order to see a relation between different charge-modulation patterns in Fig.~\ref{fig:rmap}, we calculate the Fourier transformation of $n_i$ defined by
\begin{align} 
n_\qq \equiv \frac{1}{N}\sum_i n_i e^{i\qq\cdot\rr_i},
\end{align}
where $\rr_i$ is the real-space coordinate of the site $i$.
In Fig.~\ref{fig:qmap}, we plot ${n_\qq}^2$ for $\nb=1.3$ in the momentum space. 
First, Fig.~\ref{fig:qmap}(a) shows a crystal structure factor calculated for $U=V=0$ and the uniform density $n_i=\nb$. This represents the Bragg spots due to the Penrose tiling itself, not to the charge modulation on it. The Bragg spots distribute densely while their intensity shows a 10-fold rotational symmetry.
Figure \ref{fig:qmap}(b) is an enlarged view of the first quadrant.
We can see strong intensity spots and `dotted lines' consisting of relatively weak intensity spots.  

When the electron hopping $t$ is introduced, the distribution of $n_i$ becomes inhomogeneous.
As we have seen in Fig.~\ref{fig:rmap}(c), however, this modulation at $U=V=0$ is rather weak. Therefore, in Fig.~\ref{fig:qmap}(c), we find only small changes from Fig.~\ref{fig:qmap}(b). The changes are visible around the spots denoted by red circles, which do not have a strong intensity in panel (b) but are prominent in panel (c). Namely, these spots are responsible for the weak charge modulation in Fig.~\ref{fig:rmap}(c). 

Under the electron-electron interactions $U=4V=2$, there appear new `clusters' of spots which show an increased intensity, as denoted by yellow circles in Fig.~\ref{fig:qmap}(d), while the other spots do not change much. These new spots are responsible for the strong charge modulation in Fig.~\ref{fig:rmap}(d) while the unchanged spots originate from the $\nb$ component at each site. The clustering indicates that this charge modulation is an overlap of many similar but slightly different repeating patterns.
We can also see that the clusters appear at various different $|\Vec{q}|$'s, keeping the 10-fold rotational symmetry. The different $|\Vec{q}|$'s mean different modulation periods in the real space. In fact, as presented in Fig.~\ref{fig:frac}, the real-space map of $n_i$ shows a self-similar structure, where the characteristic length scale is absent.
Thus, the difference between the two charge modulation patterns, Figs.~\ref{fig:rmap}(c) and \ref{fig:rmap}(d), is characterized by the appearance of specific clusters of the Bragg spots in momentum space. 

\section{Density of states}\label{sec:ldos}

\begin{figure}[tb]
\center{
\includegraphics[width=0.48\textwidth]{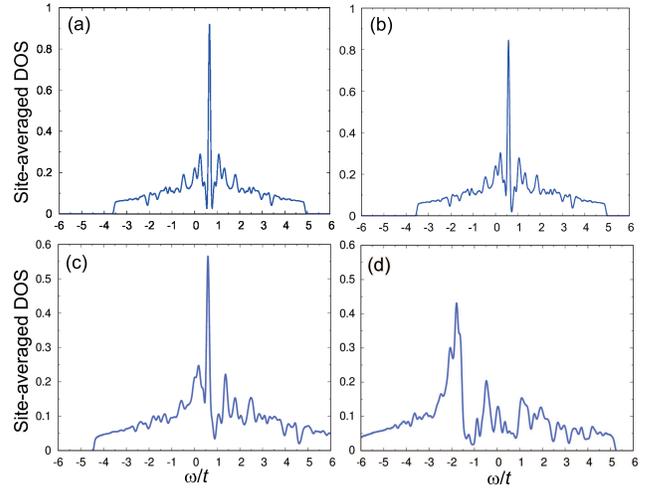}}
\caption{Site-averaged electron density of states calculated for (a) $\nb=0.7$ and $U=V=0$, (b) $\nb=0.7$ and $U=2, V=0$, (c) $\nb=0.7$ and $U=4V=2$, and (d) $\nb=1.3$ and $U=4V=2$. The calculations were done for $N=11006$. }
\label{fig:dosav}
\end{figure}

\begin{figure}[tb]
\center{
\includegraphics[width=0.48\textwidth]{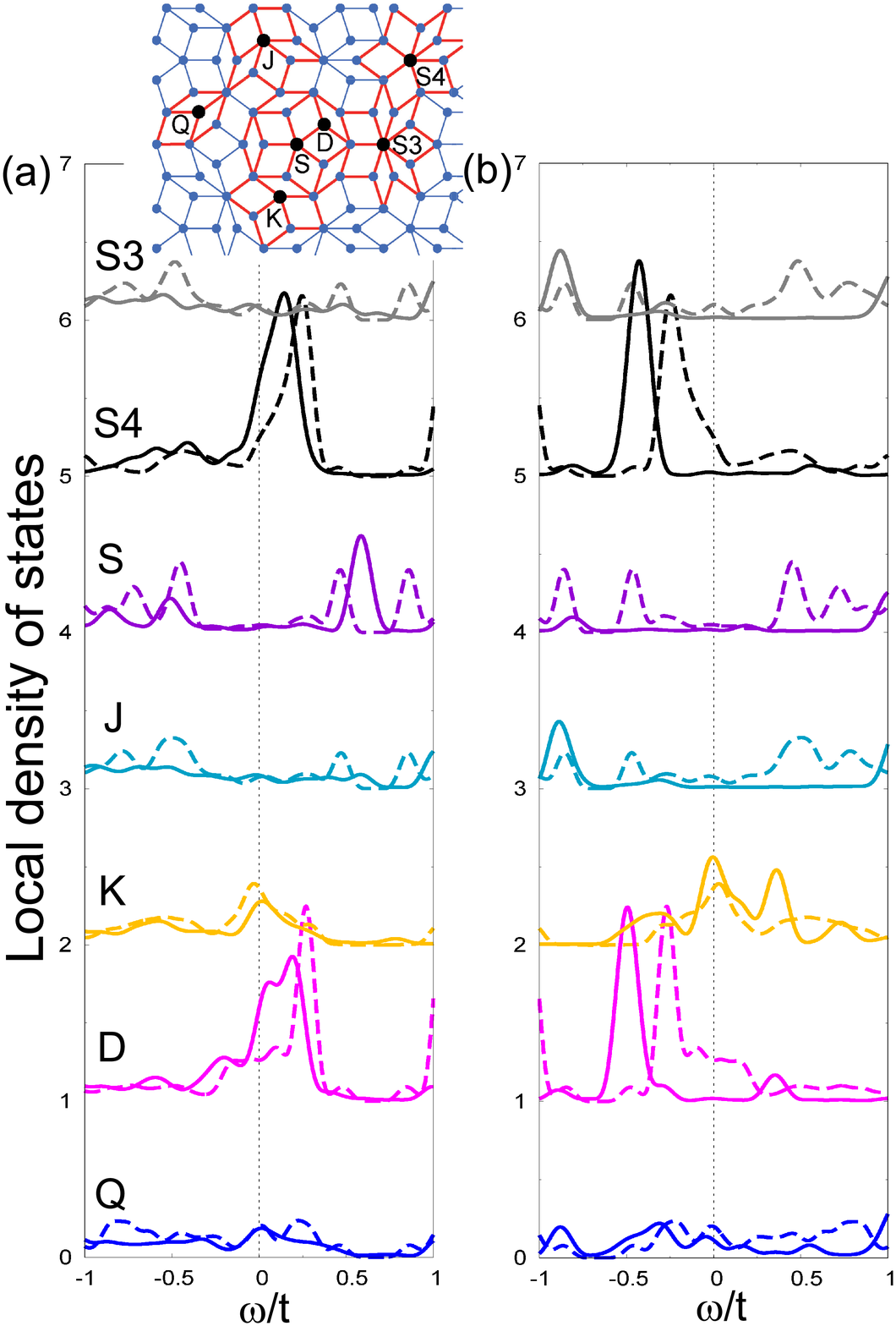}}
\caption{Local density of states at sites with various local geometries for (a) $\nb=0.7$ and (b) $\nb=1.3$. The calculations were done for $N=11006$ at $U=V=0$ (dashed curves) and $U=4V=2$ (solid curves). Each curve is shifted by 1 to the vertical direction for the sake of visibility. The inset to (a) shows the local geometry (denoted by red) of each site.  }
\label{fig:ldos}
\end{figure}

\begin{figure}[tb]
\center{
\includegraphics[width=0.48\textwidth]{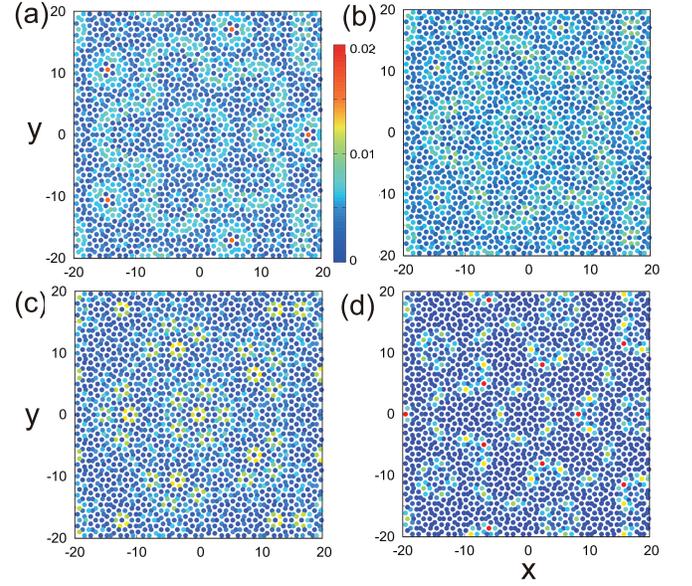}}
\caption{Real-space map of the low-energy weight of the LDOS for (a) $\nb=0.7$ (or equivalently $\nb=1.3$) and $U=V=0$, (b) $\nb=0.7$ (or equivalently $\nb=1.3$),  $U=2$ and $V=0$, (c) $\nb=0.7$ and $U=4V=2$, and (d) $\nb=1.3$ and $U=4V=2$. The calculations were done for $N=11006$. The weight is calculated by integrating the LDOS over $-0.01t <\w <0.01t$. The color bar is shared by the four panels. }
\label{fig:ldos0}
\end{figure}

Real-space modulations can also be seen in the local density of states (DOS), which is measurable by scanning-tunneling microscopy/spectroscopy (STM/STS). In fact, a recent experiment \cite{collins17} realized a STM/STS measurement of the site-dependent density of states for an artificially Penrose-tiled molecule on Cu(111) surface. 
 
To calculate the DOS in the kernel polynomial method, we used the Jackson kernel with the expansion order 500 \cite{weisse06}. 
Figure \ref{fig:dosav} shows the site-averaged DOS. 
For $U=V=0$ [panel (a)], the DOS shows a strong peak slightly above the Fermi energy. 
This peak is attributed to the confined states \cite{kohmoto86PRL,arai88}.
The DOS for $U=2$ and $V=0$ [panel (b)] is similar to that of the non-interacting system although the site-dependent mean-field gives a small difference.
Note that, as far as $V=0$, the DOS at $\nb=1.3$ is equivalent to that at $\nb=0.7$ under the electron-hole transformation.

For $V>0$, the DOS changes substantially [Figs.~\ref{fig:dosav}(c) and (d)]. 
In accord with the results in Sec.~\ref{ssec:v}, the effect of $V$ is more significant for $\nb>1$ than for $\nb<1$. Also, the DOS at the Fermi energy ($\w=0$) is more strongly suppressed for $\nb>1$ than $\nb<1$.
For $\nb=0.7$, a remnant of the confined-state peak is still prominent above the Fermi energy and a pseudogaplike behavior is seen just above it. 
On the other hand, for $\nb=1.3$, the confined-state peak seems to be strongly dampened, leaving only a weak broad peak around the Fermi energy, and the DOS seems to be suppressed for both below and above that broad peak.

Figure \ref{fig:ldos} shows the local density of states (LDOS) at various sites with different local geometries, where we follow the nomenclature by de Bruijn \cite{bruijn81-1,bruijn81-2}.
The coordination number is 3 at the sites D and Q, 4 at K, 5 at J and S, 6 at S4, and 7 at S3. 
We have chosen the sites in the central region of the 11006-site cluster so as to minimize the boundary effect.

For $\nb=0.7$, the sites D, K, and S4 show a large spectral weight at low energy while other types of sites show only small spectral weight.
The low-energy weight at sites D and S4 is even enhanced by the interaction effect while that at site K is slightly suppressed.
At $U=V=0$, the spectra at $\nb=1.3$ is the reverse (w.r.t. $\w$) of that at $\nb=0.7$.
The interaction effect, however, differs between the two.
The low-energy weight at sites D and S4 is suppressed by the interaction while that at site K is enhanced.

By integrating the LDOS in the range $-0.01t<\w<0.01t$, we define the low-energy spectral weight at each site and plot the quantity in the real space, as shown in Fig.~\ref{fig:ldos0}. We see the nonuniform pattern already in the non-interacting limit [panel (a)]. Note that, because of the electron-hole symmetry, the patterns at $\nb=0.7$ and $\nb=1.3$ are identical as far as $V=0$.  
The spatial pattern changes with the interaction [panels (b)-(d)]. 
In particular, the effect of $V$ gives a difference in the spatial patterns between $\nb=0.7$ and $\nb=1.3$.
These patterns show a self-similar structure reflecting the underlying Penrose-tiling structure, similarly to the plot of $n_i$ in Fig.~\ref{fig:frac}.

\section{Hyperuniformity analysis of the Aubry-Andr\'e-Harper model}\label{sec:aah}
\begin{figure}[tb]
\center{
\includegraphics[width=0.48\textwidth]{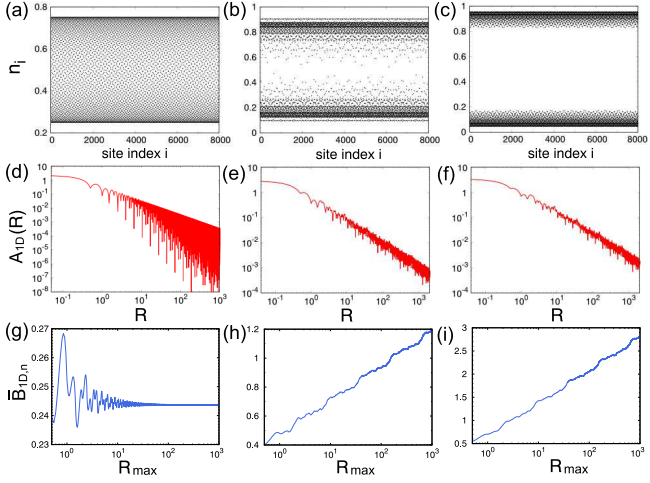}}
\caption{(a,b,c) Density distribution of the AAH model for $\lambda=t$, $2t$, and $3t$, respectively. The calculation was done for 75~025 sites, and a part of the distribution is presented for the visibility. (d,e,f) and (g,h,i) show corresponding $A_\mathrm{1D}(R)$ and $\bar{B}_\mathrm{1D,n}$, respectively.}
\label{fig:aah_n}
\end{figure}

\begin{figure}[tb]
\center{
\includegraphics[width=0.48\textwidth]{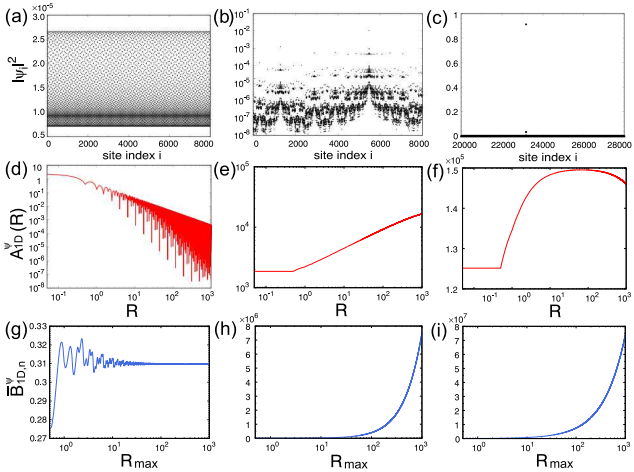}}
\caption{(a,b,c) Distribution of $|\psi_i|^2$, calculated for the ground state of the AAH model for $\lambda=t$, $2t$, and $3t$, respectively. (d,e,f) and (g,h,i) show corresponding $A_\mathrm{1D}^{\psi}(R)$ and  $\bar{B}_\mathrm{1D,n}^{\psi}$, respectively. }
\label{fig:aah_psi}
\end{figure}

Here, we study the Aubry-Andr\'e-Harper (AAH) model \cite{aubry80,harper55} in terms of hyperuniformity.
The Hamiltonian reads
\begin{align}
 H=-t\sum_i \left(c_{i+1}^\dagger c_i+c_{i}^\dagger c_{i+1}\right)+\lambda \sum_i \cos(\frac{2\pi i}{\t})c_{i}^\dagger c_i,\label{eq:aah}
\end{align}
where we consider a one-dimensional chain with the lattice constant $a=1$. $\lambda$ represents the strength of the quasiperiodic potential.
This model is known to be self-dual at $\lambda=2t$: 
The form of the Hamiltonian does not change when the basis is transformed to the momentum space.
As a consequence, for $\lambda<2t$ ($\lambda>2t$) the eigenfunction is extended (localized) in the real space.

We numerically diagonalize the chain of $N=F_n$ sites with a periodic boundary condition, where we approximate $\t$ in Eq.~(\ref{eq:aah}) to $\frac{F_n}{F_{n-1}}$ with the Fibonacci number $F_n$. 
We then calculate
\begin{align}
    A_\mathrm{1D}(R)&=\frac{2}{N}\left[\sum_{i,j}\alpha_{1D}(|i-j|;R)n_in_j -2R\sum_i n_i^2\right],\label{eq:AR1D}\\
    B_\mathrm{1D}(R)&=\frac{2R}{N}\left[\sum_{i,j}\alpha_{1D}(|i-j|;R)n_in_j -2R\sum_i n_i^2\right],\label{eq:BR1D}
\end{align}
with 
\begin{align}
  \alpha_{1D}(r;R)=\left(1-\frac{r}{2R}\right)\Theta(2R-r).
\end{align}
Equations ~(\ref{eq:AR1D}) and (\ref{eq:BR1D}) are the one-dimensional correspondence to Eqs.~(\ref{eq:AR}) and (\ref{eq:BR}), respectively.
With Eq.~(\ref{eq:BR1D}), we define the average $\bar{B}_\mathrm{1D}$ in the same way as Eq.~(\ref{eq:Bbar}).

Figures \ref{fig:aah_n}(a), (b), and (c) show the distribution of $n_i$ for $\lambda=t$, $2t$, and $3t$, where the eigenstates are extended, critical, and localized, respectively. The calculation was done for $N=F_{24}=75025$. 
The calculated $A_\mathrm{1D}(R)$ and $\bar{B}_{\mathrm{1D,n}}$ are plotted in the lower panels, where $\bar{B}_{\mathrm{1D,n}}\equiv\bar{B}_{\mathrm{1D}}/\nb^2$ with $\nb=0.5$ is the normalized value.
We find that in all the cases, $A_\mathrm{1D}(R)$ goes to zero as $R$ increases while its decay becomes slower for a larger $\lambda$ [panels (d), (e), and (f)].
Hence, these distributions are all hyperuniform.
The slow convergence is ascribed to the large fluctuation of $n_i$ as seen in Fig.~\ref{fig:aah_n}(c).

The order metric, $\bar{B}_{\mathrm{1D,n}}$, converges to 0.244 for $\lambda=t$ [Fig.~\ref{fig:aah_n}(g)]. 
This value is significantly larger than the order metric, $\frac{1}{6}$, of the point set on the integer lattice (i.e., the case of $n_i\equiv 1$) \cite{torquato03}.
On the other hand, $\bar{B}_\mathrm{1D,n}$ logarithmically increases with $R_\mathrm{max}$ for $\lambda=2t$ and $3t$ [panels (h) and (i)].
This means that these distributions are Class-II hyperuniform  \cite{torquato18}.
We have confirmed that the distribution is Class-I hyperuniform (i.e., $\bar{B}_\mathrm{1D,n}$ is constant at large $R_\mathrm{max}$) for $\lambda=1.9t$, so that the change from Class I to Class II seems to occur at the self-dual point $\lambda=2t$.

To gain a deeper insight, we also study the distribution of the ground-state wave function $\psi$.
Replacing $n_i$ in Eqs.~(\ref{eq:AR1D}) and (\ref{eq:BR1D}) with $|\psi_i|^2$, we define $ A_\mathrm{1D}^{\psi}(R)$ and $ B_\mathrm{1D}^{\psi}(R)$, and accordingly $\bar{B}_\mathrm{1D,n}^{\psi}$.
As mentioned above, $\psi$ is extended (localized) for $\lambda<2t$ ($\lambda>2t$) and critical at $\lambda=2t$.

Figures \ref{fig:aah_psi}(a), (b), and (c) present the distribution of $|\psi_i|^2$ in these three states.
When $\psi$ is extended, $A_\mathrm{1D}^{\psi}(R)$ goes to zero as $R$ increases [Fig.~\ref{fig:aah_psi}(d)].
In this case, the distribution is hyperuniform and 
the order metric $\bar{B}_\mathrm{1D,n}^{\psi}$ is well defined [Fig.~\ref{fig:aah_psi}(g)].

When $\psi$ is critical, $A_\mathrm{1D}^{\psi}(R)$ increases with $R$ [Fig.~\ref{fig:aah_psi}(e)], reflecting the self-similar distribution of $|\psi_i|^2$.
Therefore, the distribution is not hyperuniform.

When $\psi$ is localized, $A_\mathrm{1D}^{\psi}(R)$ first increases with $R$ and peaks at some value [Fig.~\ref{fig:aah_psi}(f)].
In this case, $\psi_i$ is zero almost everywhere except for a small area, so that it is not very meaningful to consider the variance, which depends on the system size.
Hence, the distribution is obviously nonhyperuniform.

To summarize, we find that the density distribution is Class-I hyperuniform for $\lambda<2t$ and Class-II hyperuniform for $\lambda\geq 2t$. On the other hand, the distribution of the wave-function amplitude is hyperuniform for $\lambda<2t$ and nonhyperuniform for $\lambda\geq2t$.

\end{document}